%% file: vp5.tex
\documentstyle[times]{mn}
\input {epsf_mn.tex}

\def\figinsert#1#2{\epsfbox{#1} \message{#2} }
\begin{document}

\def\propsima{$\; \buildrel \sim \over \propto   \;$}
\def\propsim{\lower.5ex\hbox{\propsima}}

\title
[The X-ray variability of high redshift QSOs]
{The X-ray variability of high redshift QSOs}
\author[J. Manners et al. ]
{J. Manners,
O.~Almaini,
A. Lawrence 
\\
Institute for Astronomy, University of Edinburgh, 
Royal Observatory, Blackford Hill, Edinburgh EH9 3HJ}
\date{MNRAS accepted 22 October 2001}
\maketitle

\begin{abstract}
We present an analysis of X-ray variability in a sample of 156 radio
quiet quasars taken from the ROSAT archive, covering a redshift range
$0.1 < z < 4.1$. A maximum likelihood method is used to constrain
the amplitude of variability in these low signal to noise light
curves. Through combining these data in ensembles we are able to
identify trends in variability amplitude with luminosity and with
redshift. The decline in variability amplitude with luminosity
identified in local AGN ($z < 0.1$) is confirmed out to $z=2$. There
is tentative evidence for an increase in QSO X-ray variability amplitude
towards high redshifts ($z > 2$) in the sense that QSOs of the same
X-ray luminosity are more variable at $z > 2$. We discuss possible
explanations for this effect. The simplest explanation may be that
high redshift QSOs are accreting at a larger fraction of the Eddington
limit than local AGN. 
\end{abstract}

\section{Introduction}

Rapid X-ray variability  appears to be very common in AGN (e.g. Turner
1988). Light curves can often look qualitatively
very different, but they generally appear completely random in nature with
no characteristic time-scales or periodicities, indicating there are
no long-lived orbiting components. Power spectra show `red noise'
(i.e. more power at lower frequencies), with the form $P(f) \propto
f^{-\alpha}$ where $\alpha \approx 1.5$ (Lawrence \& Papadakis 1993,
Green et al 1993). Departures from a featureless power spectrum are
rare. Some evidence for quasi-periodic oscillations has been observed
in NGC 5548, NGC 4051 (Papadakis \& Lawrence 1993, 1995) and IRAS
18325-5926 (Iwasawa et al 1998), and in a handful of AGN a turnover
has been seen at low frequencies (e.g. Edelson \& Nandra 1999). A high
frequency cut-off would indicate the size of 
the emission region, although this cannot yet be distinguished from
the noise for even the most well studied AGN. Simulated light curves
based on the existence of a number of independent flaring regions
(i.e. exponential shots) can reproduce the shape of the power spectrum
only when the shots are allowed to vary in time-scale. Beyond this, models
are difficult to constrain (Green et al 1993).

Variability studies of local AGN $(z < 0.1)$ indicate that more
luminous sources vary with a lower amplitude. This may be explained if 
more luminous sources are physically larger in size, so that they are
actually varying more slowly. Alternatively, they may contain more 
independently flaring
regions and so have a genuinely lower amplitude. The slope of this
correlation has been calculated in a number of papers using
overlapping samples of local AGN. Lawrence \& Papadakis (1993) and
Green et al (1993) analyzed samples of light curves from the EXOSAT
database. The variability amplitude was found to vary with luminosity
as $\sigma \propto L_{X}^{-\beta}$ with $\beta \approx 0.3$. The most
comprehensive analysis of the variability-luminosity relation was
carried out
by Nandra et al, 1997 (hereafter N97) for 18 local Seyferts observed
with the ASCA satellite. They find $\beta = 0.355 \pm 0.015$. Whether
this well-defined correlation applies to high redshift QSOs is not so
clear. Observations of distant QSOs are generally
of low signal-to-noise and measurements of variability in individual
objects are poorly defined. Almaini et al (2000) developed a technique 
to measure the amplitude of variability for low signal-to-noise
sources and thus high-redshift AGN. By combining light curves from a
number of AGN they were able to measure the amplitude of
variability over ranges in luminosity and redshift. They studied a
sample of 86 QSOs from the Deep ROSAT Survey of Shanks et al (1991) spanning
a wide range in redshift $(0.1 < z < 3.2)$. The behaviour of variability
amplitude with luminosity was found to be in rough agreement with the
anti-correlation seen in local AGN but showing a possible upturn for
the most luminous sources. Tentative evidence suggested this was due
to increased variability at high redshifts, although a definite trend
in the redshift behaviour could not be clearly confirmed.

In this paper we use the techniques of Almaini et al (2000) to
determine the amplitude of variability in an expanded sample of QSOs
taken from the ROSAT archive. QSOs at $z > 1$ are
preferentially selected in order to constrain the redshift behaviour
of X-ray variability. A cosmology with $q_{0} = 0.5$, $H_{0} = 50$ km
s$^{-1}$ Mpc$^{-1}$ is used throughout.

\begin{table*}
\begin {center}
\caption{Samples selected from the ROSAT PSPC archive.}
\begin {tabular}{||c|c|c||}
{\bf Subsample} & {\bf Notes} & {\bf No. QSOs} \\
\hline
Deep ROSAT Survey & 7 deep ROSAT pointings with
exposures from 30 - 80ks over 2-14 days. Broad & 84 \\
(Shanks et al. 1991)& emission line QSOs, FWHM $>$
1000 km s$^{-1}$ (Detailed in Almaini et al. 2000.)&\\
\hline
ROSAT Radio Quiet Quasar& Cross-correlation of ROSAT archive
sources with the Veron 1993 &28\\
catalogue (LEDAS)& catalogue for radio quiet
`broad emission line' objects with  z $>$ 1&\\
\hline
ROSAT deep survey of the Lockman&Data taken from single 65ks exposure,
to measure short time-scale variations. &14\\
Hole (Hasinger et al. 1998)& Object classes a - c selected (broad line
AGN). IDs taken from Schmidt et al. 1998. &\\
\hline
Deep ROSAT Survey&Data taken from the longer first exposure of
73ks. &13\\
(McHardy et al. 1998)& Broad emission line objects selected (FWFM $>$
1000 km s$^{-1}$) &\\
\hline
Veron 2000 cross-correlated with& Acts as an update to the ROSAT RQQ
catalogue. QSOs selected with z $>$ 2, &17\\
ROSAT archive sources (LEDAS)& undetected in radio. Veron QSOs are
defined to have `broad emission lines'. &\\
\hline
&{\bf \hspace{7cm} Total:}&{\bf 156}
\end{tabular}
\end{center}
\end{table*}

\section{The sample}

The sample consists of 156 QSOs between $0.08 < z < 4.11$ taken from
the ROSAT PSPC archive. It is made up of QSOs taken from a number of
sources that all adhere to the following selection criteria:

\begin{itemize}
\item ID: radio quiet quasar
\item X-ray exposure $>$ 10,000 seconds
\item Flux signal-to-noise $>$ 5
\item Within 20 arcmin of ROSAT pointing
\item Optimal re-binning (see section 3) gives at least 3 time bins
\end{itemize}

The sample can be broken down into 5 subsamples, individually selected
to cover the luminosity and redshift parameter space. These are listed
in Table 1. Only PSPC data was used for reasons of consistency and
ease of data reduction. The redshift distribution of the entire sample is
displayed in Fig. 1.

\begin{figure}
\caption{Redshift distribution of our sample of 156 radio quiet quasars.}  
\centering \centerline{\epsfxsize=8.5 truecm
\figinsert{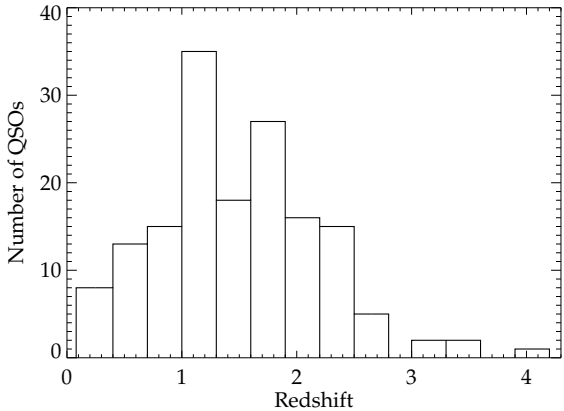}{0.0pt}}
\end{figure}

\section{Data reduction}

The data were obtained from the LEDAS online database facility at
Leicester. Data reduction was performed with the Asterix X-ray data
processing package. Each source was extracted using a circular mask of
radius chosen to include 90 per cent of the PSF. The data was then filtered
to remove periods of high particle background characterized by the
Master Veto Rate rising above 170 counts s$^{-1}$. Additional filtering
restricted the energy range to 0.1 -- 2.4 keV.

Careful background subtraction was extremely important to ensure
measurement of source variability was not compromised. Background
regions were chosen to be as close as possible to the source to
minimize the effects of any background gradients. Areas of at least 5
arcmin radius were used. These were large enough to smear out
background irregularities due to undetected sources. A number of
further regions were selected and compared with the first to ensure that
our chosen region did not contain any abnormalities. Finally, the
background light curve was compared with the source light curve and a
linear correlation coefficient computed. If a highly significant
correlation (or anti-correlation) was found the data was reduced again
using a different background region. 

Binning of the light curves was initially constrained by the orbit of
the satellite. ROSAT's low altitude (580km), and overheads led to a
typical exposure of 1000-2000 seconds per 96 minute orbit. The light
curves were therefore binned on these periodic orbits. An algorithm
was then used to bin up the data to allow meaningful Gaussian
statistics. Re-binning of light curves can be approached in a number
of ways, and important data may be lost if the method used is
over-simplified. The algorithm constructed uses the mean intensity of
the source to identify bins that would nominally contain less than 15
photons. These bins are merged with the neighbouring bin that has been
merged the least number of times. Where large gaps in the data are
identified ($>$ 5 hours), bins will be discarded rather than merged
with data many orbits away. Light curves with less than 3 time bins
were not used for the variability analysis.

In order to remain consistent, all measurements of X-ray luminosity
are made through extraction of fluxes from the same datasets used
for the variability analysis.

\section{Measuring variability}

The method used for measuring the amplitude of intrinsic variability
is fully described in Almaini et al (2000). To compare objects of
different flux, the light curves are divided by the mean, so that
measurements of {\em fractional} variability are made. A maximum likelihood
technique is used to separate the intrinsic variations in the light
curve from those due to noise. The likelihood for values of QSO
variability amplitude ($\sigma_Q$) is given by:

\begin{equation}
L(\sigma_Q|x_i,\sigma_i)=\prod_{i=1}^N\frac{\exp\left\{-\frac{1}{2}
                         (x_i-\bar{x})^2/(\sigma_i^2+\sigma_Q^2)\right\}}
                         {(2\pi)^{1/2}(\sigma_i^2+\sigma_Q^2)^{1/2}}
\end{equation}

\noindent where $x_i$ are data points with mean $\bar{x}$ (1 in this case) and
$\sigma_i$ are the measurement errors. The maximum likelihood estimate
for $\sigma_Q$ is obtained from the peak of the distribution. The
errors are measured by finding limits of equal likelihood that enclose
68 per cent of the area under the likelihood curve.

\begin{figure}
\caption{(a) Maximum likelihood estimates for the variability
amplitude as a function of luminosity for the 156 QSOs. The size of
the points indicate relative flux. In (b) we
display the results in ensemble form (the vertical scale has been
changed). Note, these are calculated from the original light curves
and are not averages of the individual likelihood estimates (see
section 4). The exclusion of QSO 0015+1603 is explained in section 6.} 

\centering
\centerline{\epsfxsize=8.5 truecm \figinsert{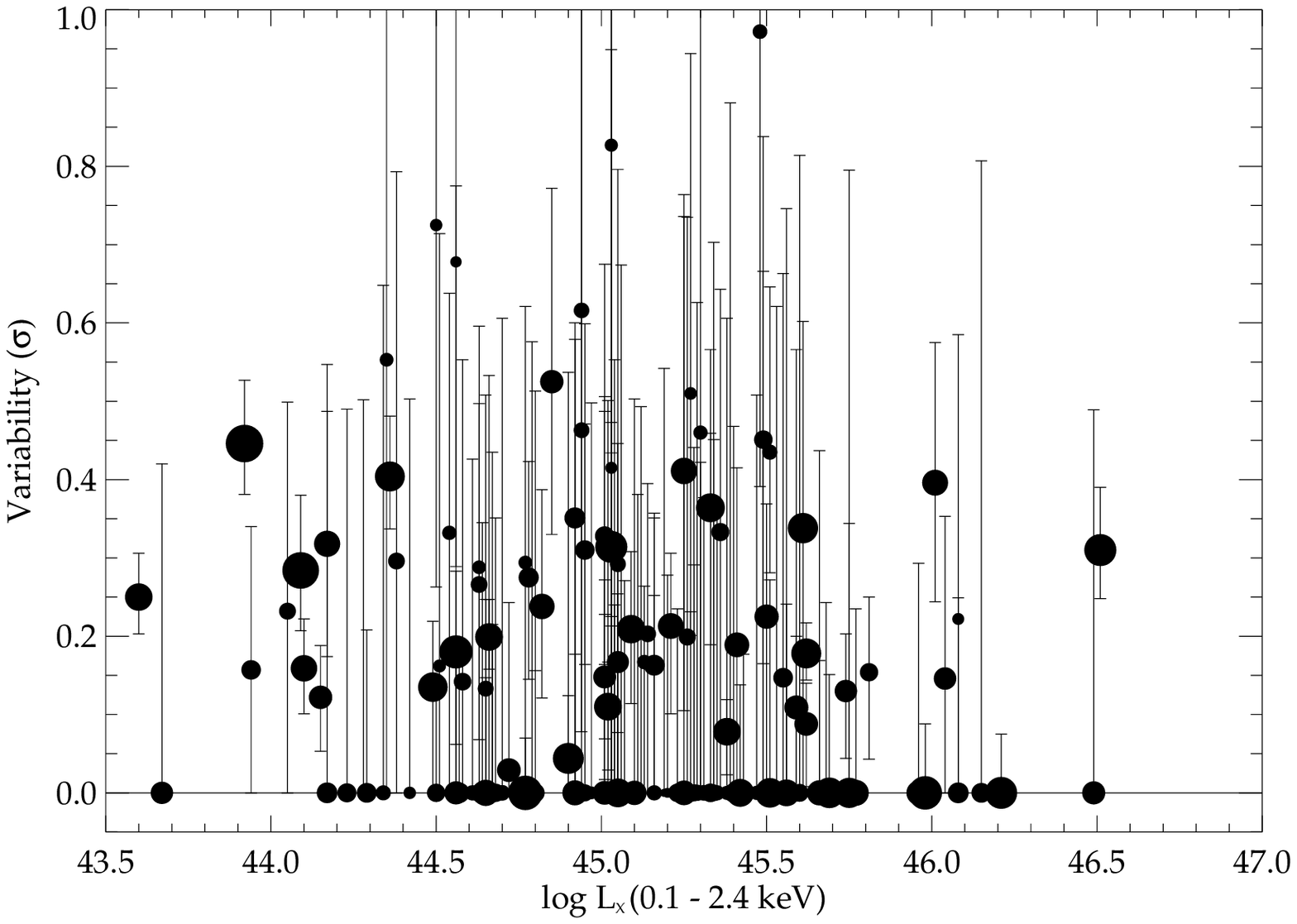}{0.0pt}}
\centerline{\epsfxsize=8.5 truecm \figinsert{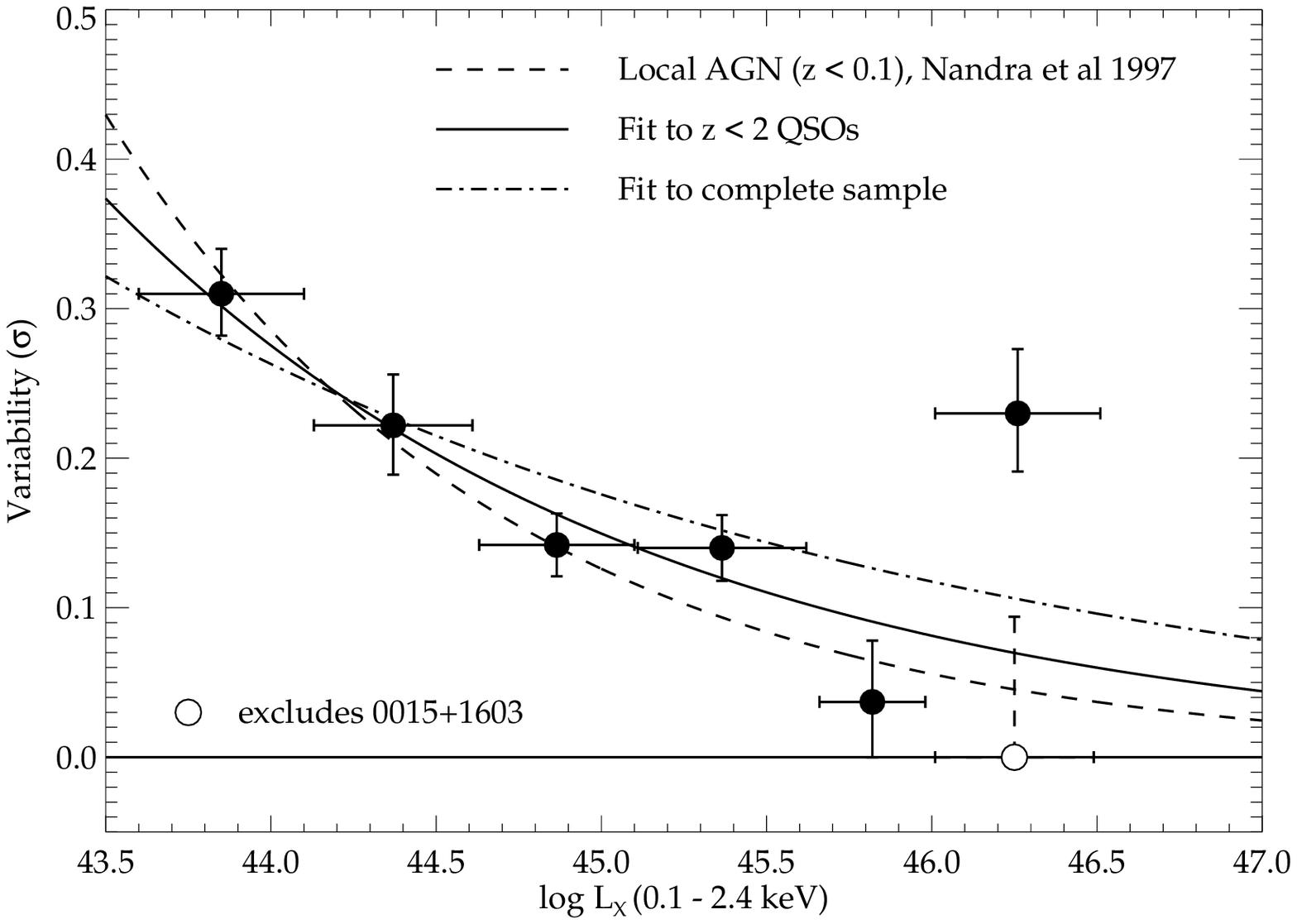}{0.0pt}}
\end{figure}

\begin{figure}
\caption{(a) Maximum likelihood estimates for the variability
amplitude as a function of redshift for the 156 QSOs. The size of
the points indicate relative flux. In (b) we display the results in
ensemble form (the vertical scale has been changed). Note, these are
calculated from the original light curves and are not averages of the
individual likelihood estimates (see section 4). The exclusion of QSO
0015+1603 is explained in section 6.}

\centering
\centerline{\epsfxsize=8.5 truecm \figinsert{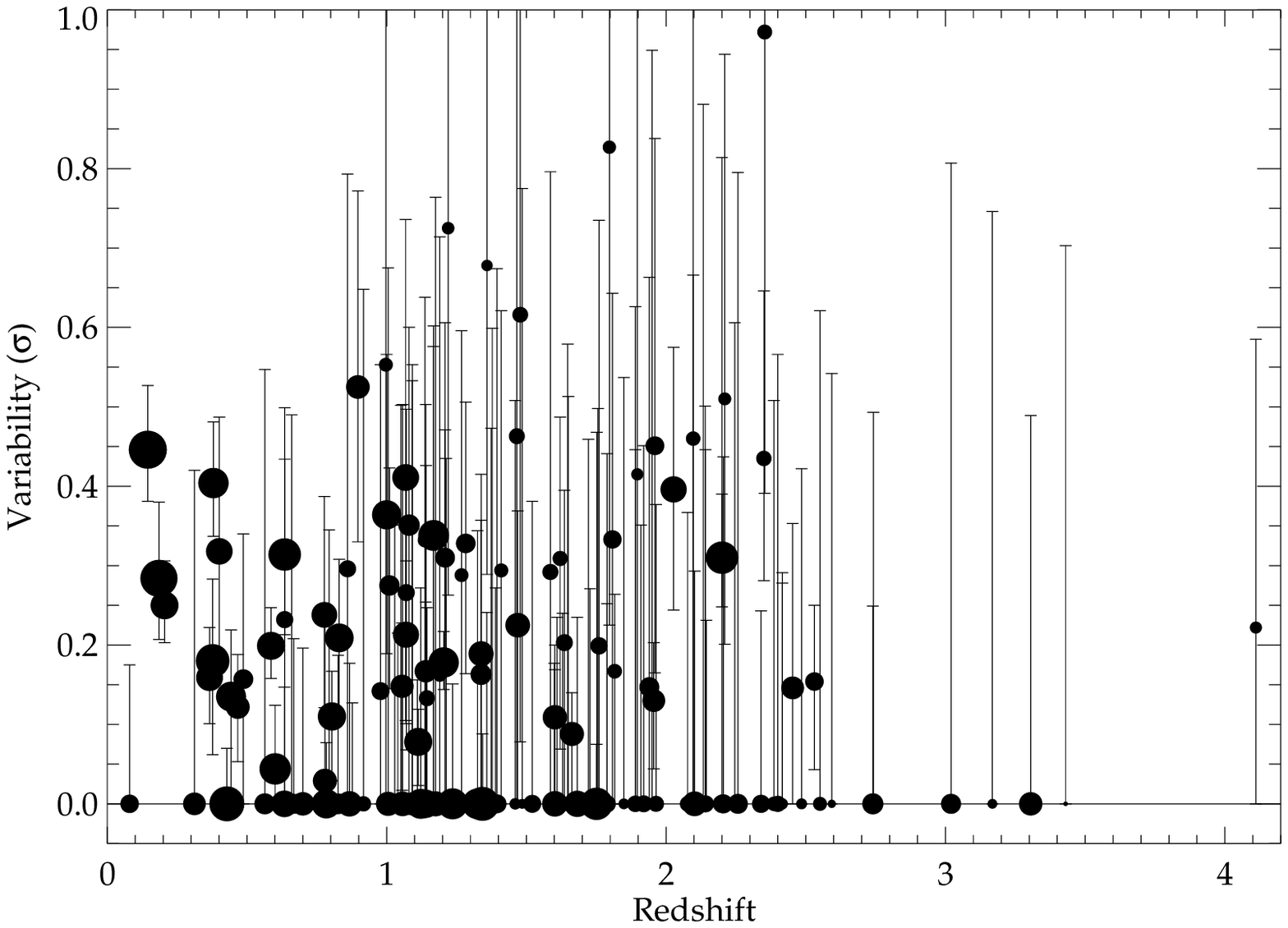}{0.0pt} }
\centerline{\epsfxsize=8.5 truecm \figinsert{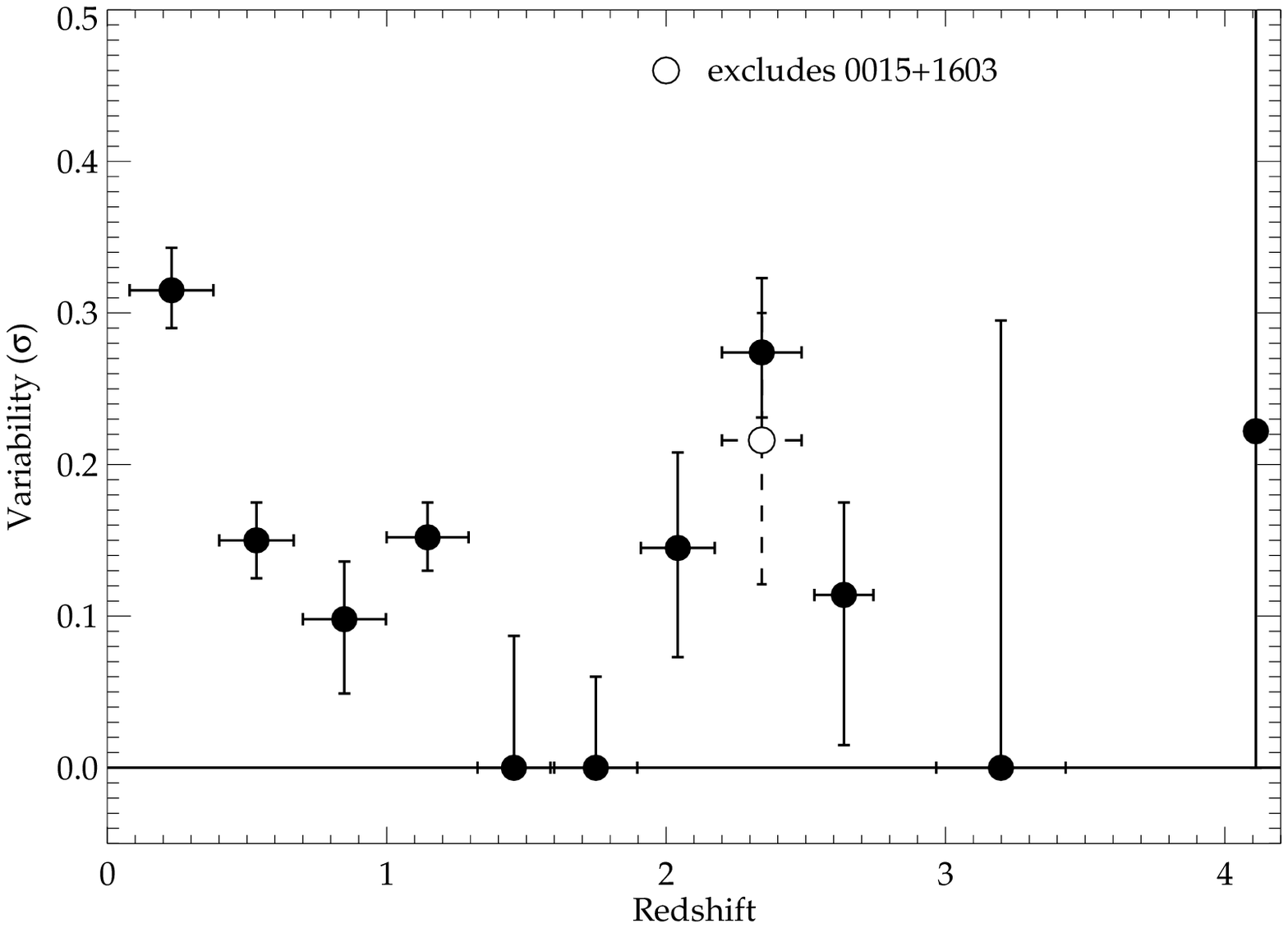}{0.0pt}}
\end{figure}

The majority of the light curves analyzed here are of low
signal-to-noise which in many cases can only give us an upper limit on
the amplitude of variability. In order to extract meaningful
information from the sample it was necessary to combine the light
curves into ensembles over given ranges of luminosity or redshift. 
To do this we assume that all the quasars in an ensemble have the same
intrinsic variability. We can then exploit the fact that variance does
not depend on the order of the measurements to effectively combine all
the light curves into one `ensemble light curve'. The amplitude of
intrinsic variability can then be measured as for a single
object. This provides a more comprehensive, accurate and unbiased
method of determining variability within an ensemble than simply
taking an average of the individual maximum likelihood estimates. Of
course, if the quasars in a given luminosity or redshift bin actually
have a range of intrinsic variabilities, our estimate may be biased
towards the most variable members. We discuss this possibility in
section 6.

A number of corrections must be made to the variability measurements
before they can be compared in an unbiased way. The  serendipitous
nature of this sample means that light curve observations have a range
of total integration times and sampling rates. QSO power spectra show
`red noise' indicating that the observed amplitude of variability will
increase with length of observation. We have therefore normalized
variability amplitudes to a time-scale of 1 week, assuming the power
spectrum slope seen in local AGN ($\alpha = 1.5$). This is also used
to correct for time dilation effects since the frequency of variations from
distant quasars will be decreased, lowering the measured amplitude of
variability for observations of finite length. 

A further correction is made for the effects of irregular binning. Low
signal-to-noise light curves tend to have longer time bins for which
high frequency variations are smeared out. The details of the
correction methods used can be found in Almaini et al (2000). 

\section{The results}

\begin{figure*}
\caption{Variability amplitude as a function of luminosity over 3
redshift ranges. The line plotted is the best fit relation to local 
($z < 0.1$) AGN found by Nandra et al (1997).}  
\centering \centerline{\epsfxsize=15.0 truecm
\figinsert{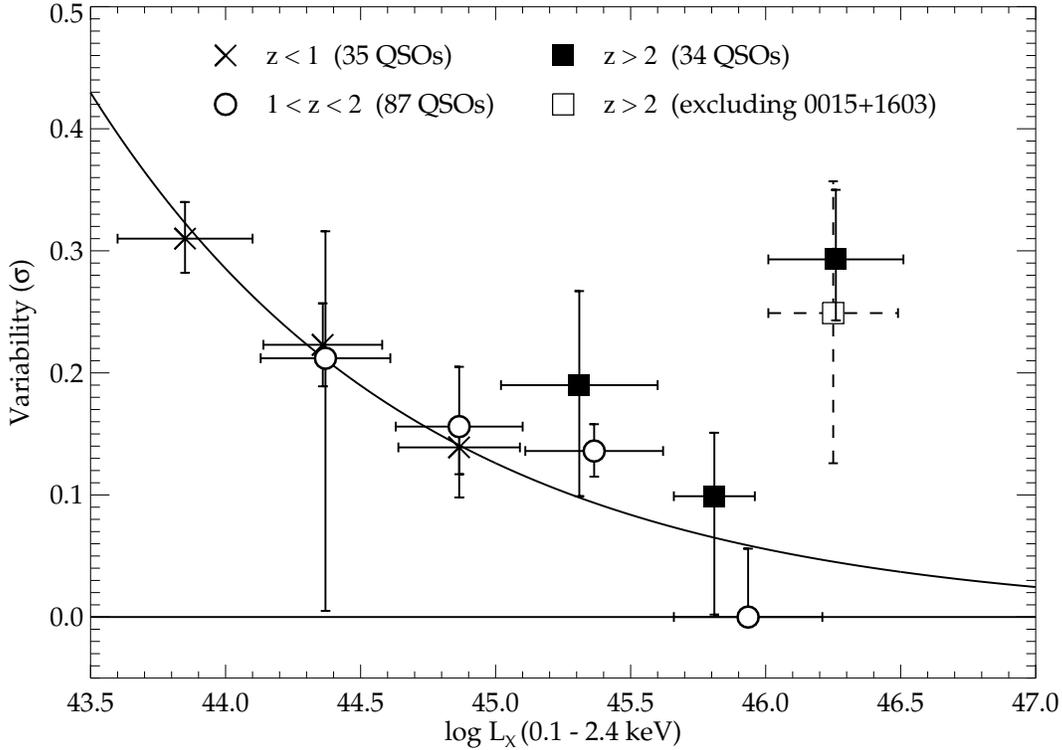}{0.0pt}}
\end{figure*}

The maximum likelihood estimates for the corrected intrinsic variability
amplitudes of all 156 quasars are displayed in Fig. 2 and 3. The
size of the points in Fig. 2(a) and 3(a) indicate relative
flux. This is done in order to distinguish objects of higher
signal-to-noise. Nearly half the sample (69 QSOs) give a non-zero
maximum likelihood estimate of variability amplitude. The remaining 
QSOs can only
provide upper limits, mainly due to low signal-to-noise. Treating the
entire sample as one ensemble, the corrected mean ensemble variability
is: $\sigma = 0.15 \pm 0.01$ ($\sigma = 0.16 \pm 0.01$ uncorrected) on
a time-scale of 1 week.

\subsection{Correlations With Redshift And Luminosity}

In Fig. 2 and 3 we display the variability amplitude as a function
of luminosity and redshift. The data are split into 6 luminosity bins
each of $\sim 0.5$ dex, and 11 redshift bins each of $\delta z = 0.3$. 
The ensemble variability amplitudes were calculated for each bin by
combining the individual light curves and performing a maximum
likelihood analysis as described in section 4. 

Immediately apparent in Fig. 2(b) is an anti-correlation between
quasar luminosity and variability amplitude. A power law fit to the
individual quasars gives the best fit relation: $\sigma \propto
L_{X}^{-\beta}$ where $\beta = 0.18 \pm 0.05$ (plotted as a dot-dash 
line). This gives a fairly poor fit to the ensemble points, mainly due
to some highly variable, high luminosity
QSOs. However, a power law fit to QSOs with redshift less than 2
(solid line, $\beta = 0.27 \pm 0.05$) passes through the majority of
the ensemble points and is very close to the average relation found
for local AGN (plotted as a dashed line). The errors quoted here are
68 per cent confidence limits for the slope on allowing the normalization to
float to its optimum value.

\begin {table}
\begin {center}
\caption {Showing agreement between redshift ensembles from Fig. 4,
and the best fit relation to local AGN found by Nandra et al ($\sigma
\propto L_{X}^{-0.355}$)} 
\begin {tabular}{||c|c|c||}
{\bf QSO sample} & $\mathbf{\chi^2_{red}}${\bf fit to local AGN} & {\bf Probability} \\
\hline
$z < 1$ & 0.1 & 87\%\\
$1 < z < 2$ & 1.7 & 17\% \\
$z > 2$ & 12.9 & 0.00001\%\\
$z > 2$ (excl. 0015+1603) & 2.0 & 14\%\\
\hline
\end{tabular}
\end{center}
\end{table}

\begin{figure}
\caption{Variability amplitude as a function of redshift after
removing the luminosity dependence (normalizing to
$L_{X}=10^{45}$). In (a) we use the best-fit power law to QSOs of $z
< 2$ of $L_{X}^{-0.27}$. In (b) the relation for local AGN from
Nandra et al (1997) of $L_{X}^{-0.355}$ is used.}
\centering \centerline{\epsfxsize=8.5 truecm
\figinsert{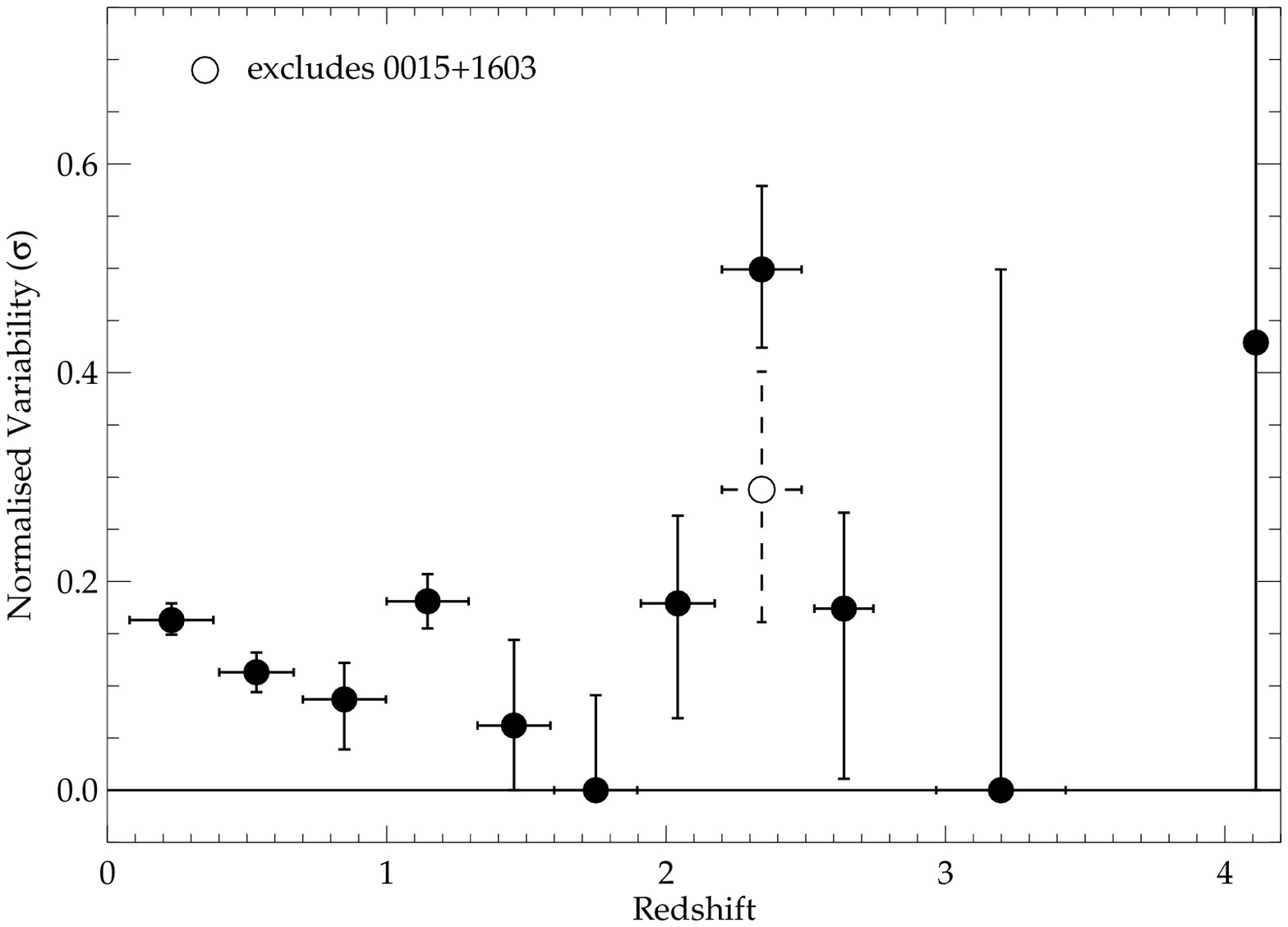}{0.0pt}} 
\centerline{\epsfxsize=8.5 truecm
\figinsert{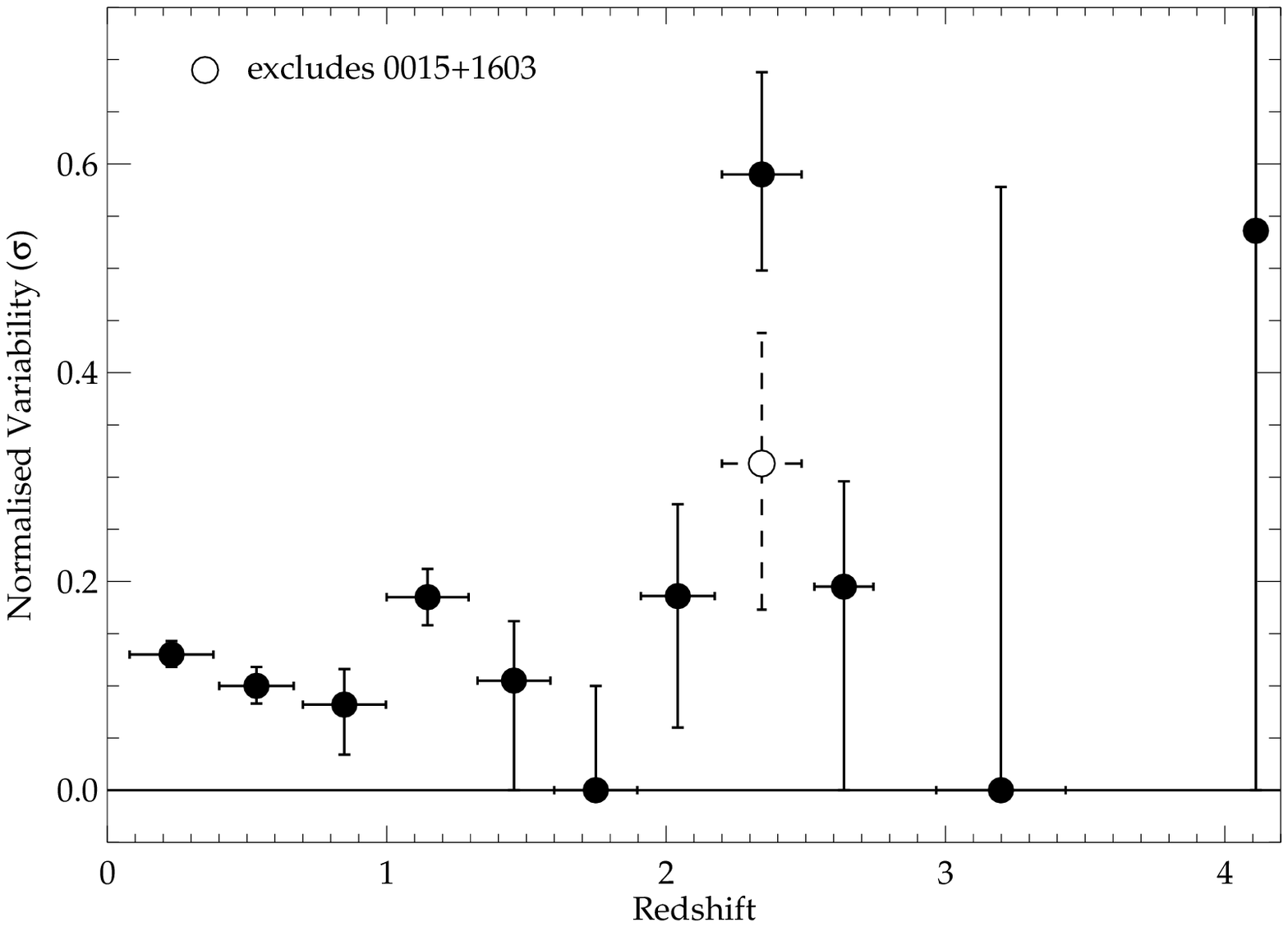}{0.0pt}}
\end{figure}

For the redshift dependence (Fig. 3b), it is reasonable to expect a
certain amount of degeneracy with luminosity. This effect appears to
dominate at low and medium redshifts. At redshifts beyond $\sim$ 2, an
upturn in variability is observed which cannot be explained as a
consequence of the known trend with luminosity.

In order to decouple the effects of luminosity and redshift, the
variability amplitude was plotted as a function of luminosity for 3
redshift intervals (Fig. 4). The luminosity bins for each redshift
interval overlap, providing a measure of the 
change in variability amplitude as a function of redshift. Comparing
the first 2 redshift intervals we find no significant difference in
the $\sigma - L_X$ relation. The anti-correlation between variability
and luminosity appears to be unchanged out to a redshift of 2. To compare
this with the relationship for local AGN, the power-law slope found by
N97 is plotted on Fig. 4. This has been normalized
to fit the points in the first redshift interval.\footnote{It would be
incorrect to use the normalization of N97, mainly due to the different
observation lengths used (typically less than one day) and the
different spectral range used to calculate luminosities. Crude
corrections based on a standard power spectrum and X-ray spectral
index give variability amplitudes around a factor of 2 lower for the
N97 AGN. Given the uncertainties in these corrections and the
differing methods of calculating variability amplitudes, this is not
thought to be significant.} The reduced $\chi^2$
values given in Table 2 compare the points in each 
redshift range with the normalized local AGN relation. Low values for
the first 2 intervals display the observed close agreement with local
AGN. The high value of reduced $\chi^2$ for $z > 2$  can be mostly
attributed to the object 0015+1603. However, once this object is
removed, the hypothesis that these quasars follow the local $\sigma -
L_{X}$ relation can still be rejected at the 85 per cent confidence
level. Considering that all 3 points also show an `excess' variability
indicates that high redshift QSOs may not be well characterized by the
variability-luminosity correlation of local AGN.

The upturn in variability seen for the highest redshift quasars in our
sample can be quantified by fitting the N97 slope to the ensemble
points in the redshift interval $z > 2$. This can then be compared to
the fit for the $z < 1$ redshift interval. We find, at a
fixed luminosity, that high-z QSOs are more variable than low-z QSOs by 
$\delta \sigma = 0.11$. An alternative interpretation is that
at a fixed variability, high-z QSOs are more luminous by a factor of
16. If we exclude the variable quasar 0015+1603 (see section 6), the
best fit requires  $\delta \sigma = 0.06$ or an increase in luminosity
by a factor of 6.

\subsection{Characterizing the redshift dependence}

In order to illustrate the redshift dependence of QSO
X-ray variability, we attempted to remove the effect of the luminosity
dependence by normalizing values to a luminosity of $L_{X}=10^{45}$
ergs s$^{-1}$. Due to the close agreement between the first
2 redshift intervals observed in Fig. 4, this was first done using
the best fit power-law to QSOs with $z < 2$ $(\sigma \propto
L_{X}^{-0.27})$. The results are plotted in Fig. 5. The low
redshift ensembles (below $z \sim 2$) in Fig. 5(a) still appear to
display a downward trend. A minimum is observed at a redshift of
$\sim$ 1.7 where the level of variability becomes consistent with
zero. In contrast, the high redshift ensembles $(z > 2)$ display an
increase in variability. In Fig. 5(b) we use the
luminosity relation found for local AGN (N97) to
normalize variability amplitudes. Here the redshift ensembles for QSOs
of $z < 1$ are noticeably flatter. Evidence for the minimum observed
in Fig. 5(a) has become less significant. However, beyond $z \sim
2$, the increase in variability is more pronounced.

\section{Properties of high-z variable QSOs}

The apparent increase in variability seen in high redshift QSOs may be
due to the inclusion of a new population of objects rather than
differences in the `typical' population. Narrow-line Seyfert 1s are
known to exhibit enhanced variability (Boller et al 1996, Leighly
1999) and could be responsible for this upturn if their high-z
equivalents were more prevalent than they are today. In order to test
this hypothesis, the identifications for the 12 QSOs that exhibit
detected variability at redshifts greater than 1.9 (i.e. the last 5
redshift intervals in Fig. 3(b) \& 5), were studied in
more detail (Fig. 6 and Table 3). All the QSOs with adequate
optical spectra were found to contain broad permitted lines
($>$ 3000 km s$^{-1}$ FWHM). In one object, 0015+1603, the optical grism
spectrum was not of sufficient quality to determine
linewidths. This object also displays a steep X-ray spectrum, so is a
(potential) NLS1 candidate. 0015+1603 was the highest luminosity object
in the sample 
with a high significance detection of variability. To determine the
effect of this steep-spectrum QSO on the characteristics of the
sample, the variability analysis was repeated with the object
removed. The affected bins are plotted as unfilled points in Fig. 1 -
4. Removing this highly variable object decreases the significance of
the upturn in variability for the high-z sample, but does not
affect the direction of the trend. 

\begin{table*}
\begin {center}
\caption{Notes on high-z variable quasars.}
\begin {tabular}{||c|c|c|c|c|c||}
{\bf Name} & {\bf RA, Dec.} & {\bf z} & {\bf log $L_{X}$} & {\bf Variability} & {\bf Notes} \\
X-ray source name& (J2000.0)&& (0.1-2.4keV) & ($\sigma$) & \\
\hline
{\bf 1118+1354} & 11 21 06.00, & 1.94 &
45.55 & 0.15 & ROSAT spectrum gives $\alpha \sim 0.3$ , Ly $\alpha$,
CIV present \\
{RX J112106.0+133825} & +13 38 25.1 & && +0.52, -0.15 & Faint emission line quasar
(Weedman 1985) \\
\hline
{\bf [HBG98] 031} & 10 53 31.80, & 1.956 & 45.74 & 0.13 & Broad SiIV,
CIV, CIII{]} (Lehmann et al 2000) \\
{RX J105331.8+572454} & +57 24 53.8 & && +0.07, -0.09 & \\
\hline
{\bf 0438-1638} & 04 40 26.478, & 1.96 & 45.49 & 0.45 & Broad Ly
$\alpha$, CIV (Osmer, Porter \& Green 1994) \\ 
{1E 0438-166} & -16 32 34.60 & && +0.39, -0.29 & \\
\hline
{\bf MS 0104.2+3153} & 01 06 58.756, & 2.027 & 46.01 &
0.40 & Broad CIV, SiIV (Gioia et al 1986), BAL quasar with \\
{2RXP J010659.1+320920} & +32 09 18.01 & && +0.18, -0.15 & component from foreground IGM (Komossa \&
Bohringer 1999) \\
\hline
{\bf SGP3X:021} & 00 54 47.29 & 2.097 & 45.30 & 0.46 & Broad Ly
$\alpha$, CIV ($>3000$km/s, Boyle et al 1990) \\
SGP 3:48 & -28 31 54.8 & & & +0.56, -0.46 & \\
\hline
{\bf 0015+1603} & 00 17 45.05, & 2.20 &
46.51 & 0.31 & Steep ROSAT spectrum ($\alpha \sim 2.8$) \\
{2RXP J001749.5+161948} & +16 19 52.6 & && +0.08, -0.06 & Ly $\alpha$, CIV present (Anderson \& Margon 1987) \\
\hline
{\bf F864X:013} & 13 43 09.2, & 2.21 & 45.27 & 0.51 & Broad CIII, CIV
($>3000$km/s, Almaini 1996) \\ 
(unpublished) & -00 22 57 &&& +0.43, -0.31 & \\
\hline
{\bf F864X:086} & 13 43 29.20, & 2.347 & 45.51 & 0.44 & ROSAT spectrum:
$\alpha$ = 0.6, Narrow Ly $\alpha$, CIV (Almaini et al 1995) \\ 
RX J1343.4+0001 & +00 01 33.0 & && +0.21, -0.15 & Broad H$\alpha$, no
H$\beta$, type 1.9 QSO (Georgantopoulos et al 1999) \\
\hline
{\bf CRSS J1415.1+1140} & 14 15 11.20 & 2.353 & 45.48 &
0.97 & Broad Ly $\alpha$, CIV (Boyle et al 1997)\\
{2RXP J141511.7+114000} & +11 40 03.0 & && +0.72, -0.58 & \\
\hline
{\bf POX 042} & 12 00 44.975, & 2.453 & 46.04 & 0.15 &
Broad CIV, SiIV,OIV{]} (Ulrich 1989)\\
{2RXP J120044.8-185952} & -18 59 45.04 & && +0.21, -0.15 &\\
\hline
{\bf 0315-5522} & 03 16 50.40, & 2.531 & 45.81 & 0.15 & Broad Ly $\alpha$, CIV, Si+OIV{]} (Zamorani et al 1999) \\
{[ZMH99] X036-04} & -55 11 09.9 &&& +0.10, -0.11 &\\
\hline
{\bf Q0000-26} & 00 03 22.909, & 4.111 & 46.08 & 0.22 & Broad Ly
$\alpha$, CIV, SiIV (Schneider et al 1989) \\
{2RXP J000322.6-260312} & -26 03 16.83 &&& +0.36, -0.22 & \\
\hline
\end{tabular}
\end{center}
\end{table*}

\begin{figure}
\caption{High-z variable QSOs.}
\centerline{\epsfxsize=8.0 truecm \figinsert{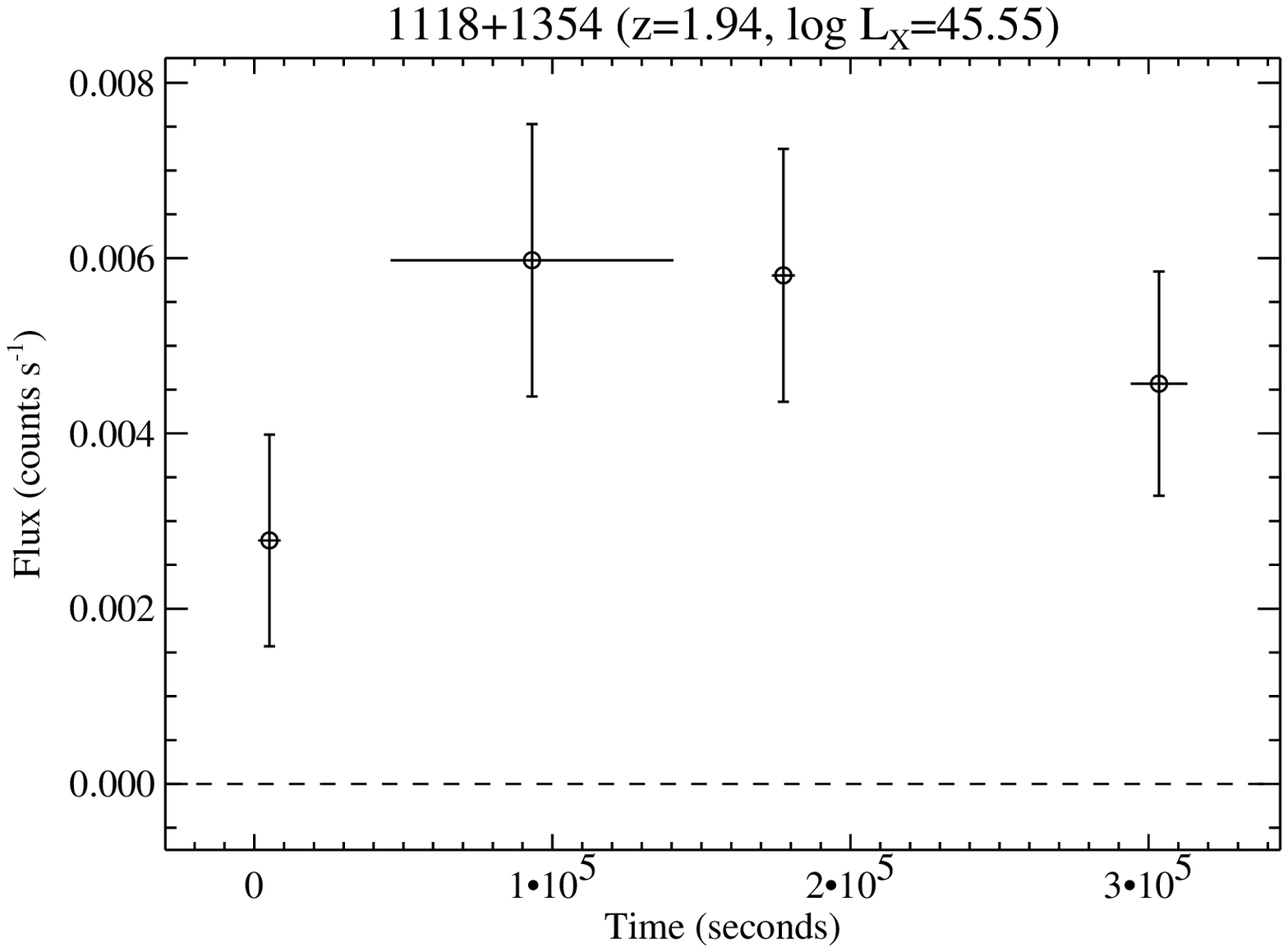}{0.0pt}}
\centerline{\epsfxsize=8.0 truecm \figinsert{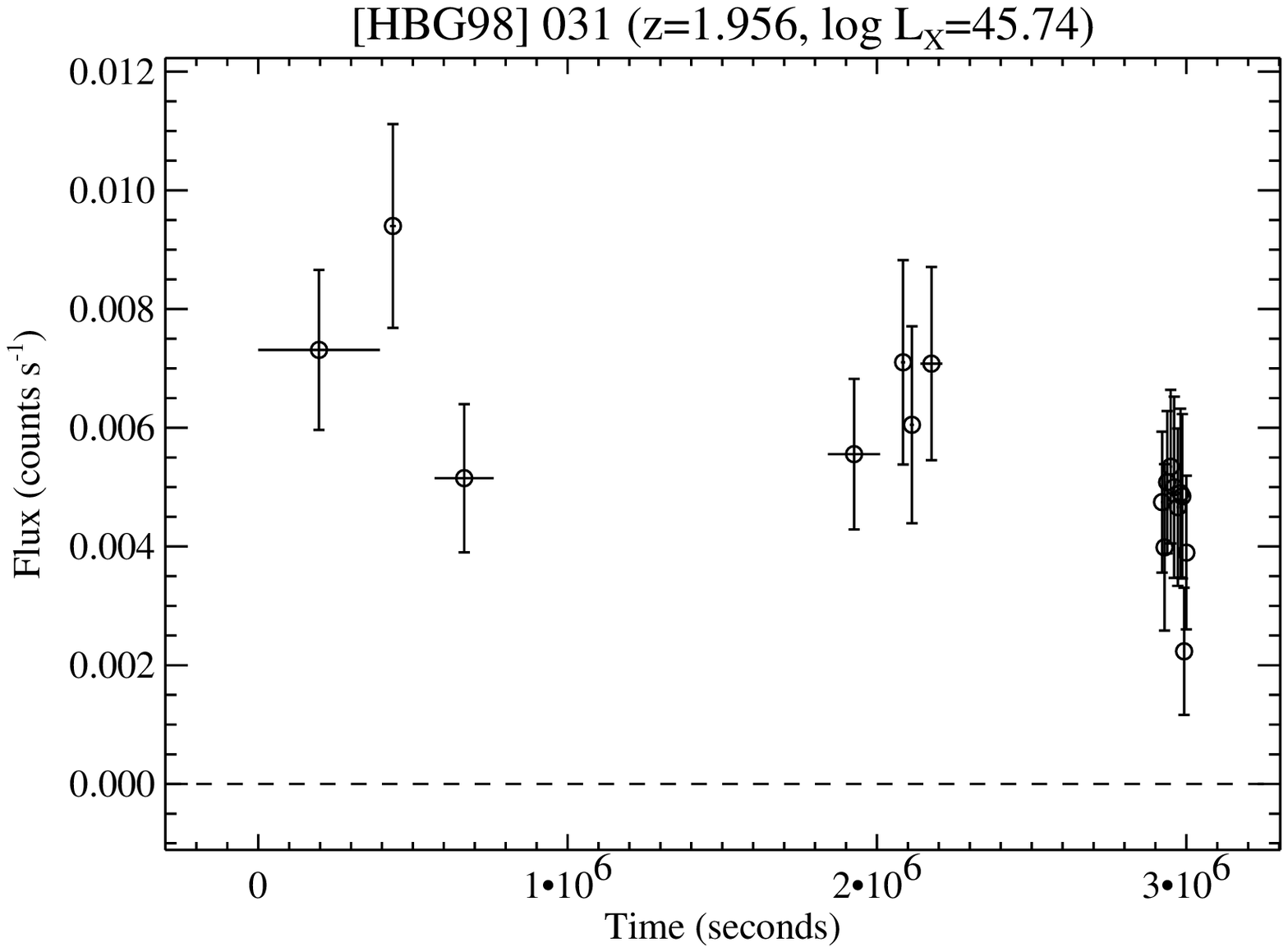}{0.0pt}}
\centerline{\epsfxsize=8.0 truecm \figinsert{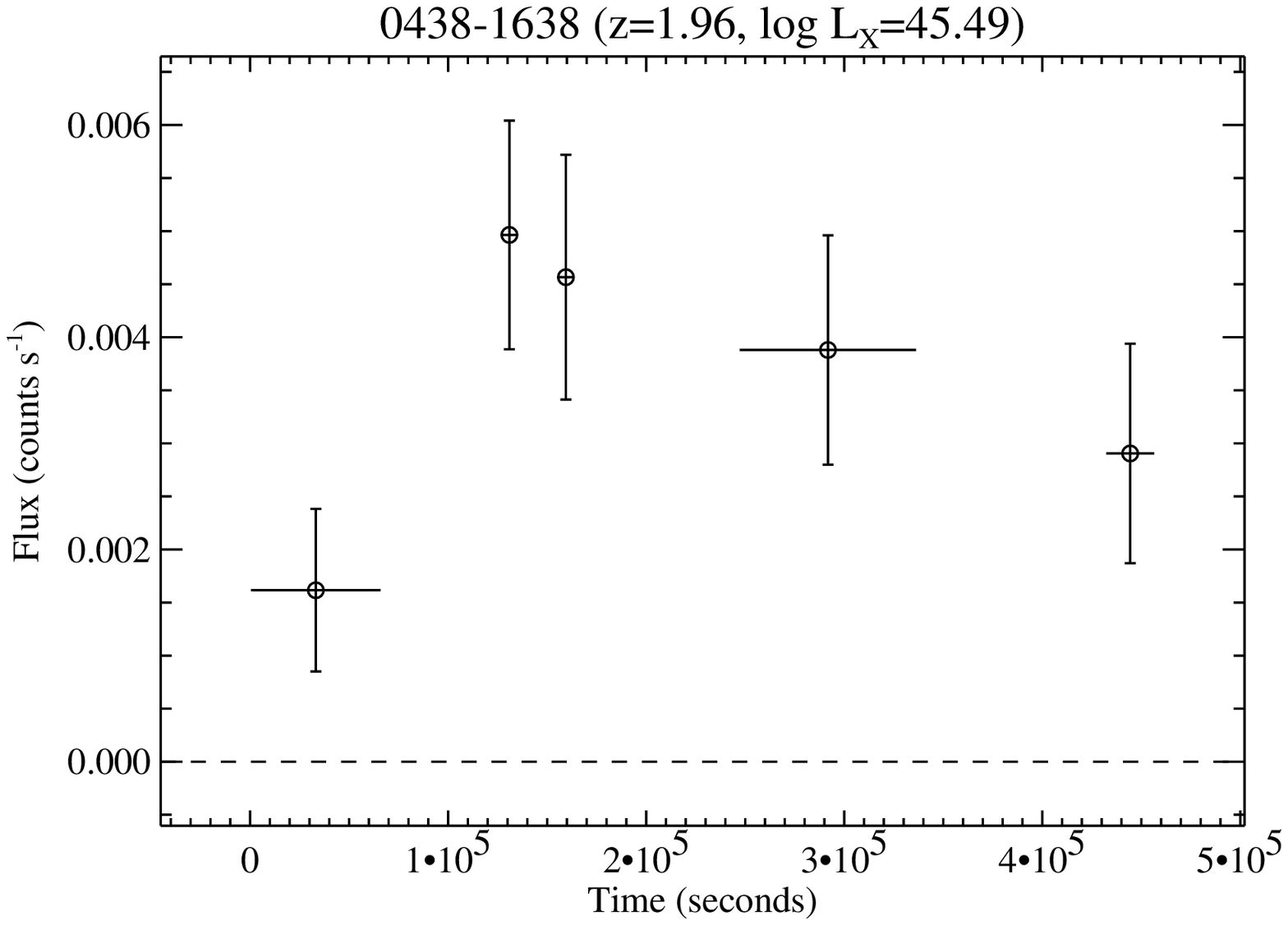}{0.0pt}}
\centerline{\epsfxsize=8.0 truecm \figinsert{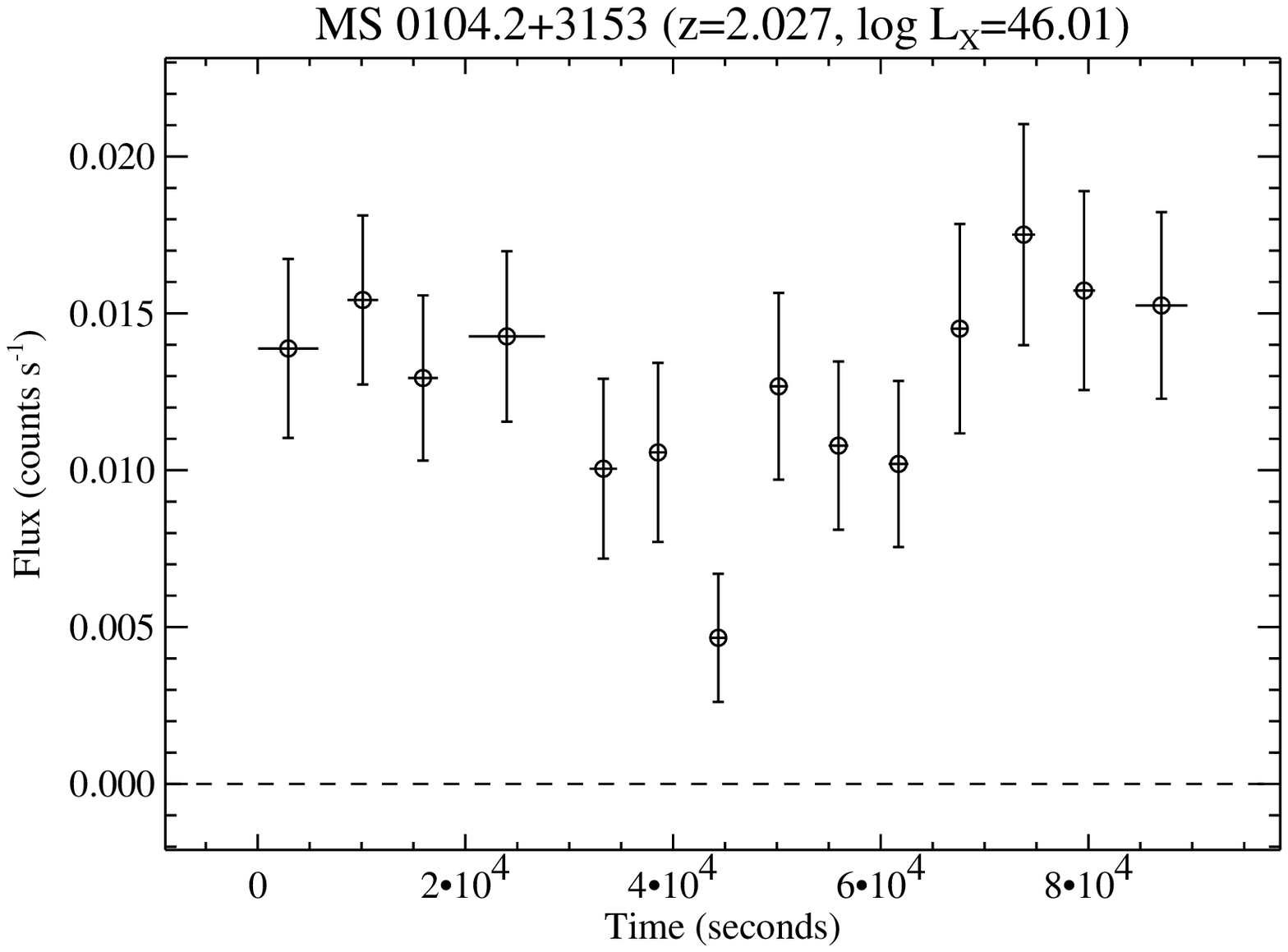}{0.0pt}}
\end{figure}
\addtocounter{figure}{-1}
\begin{figure}
\caption{High-z variable QSOs.}
\centerline{\epsfxsize=8.0 truecm \figinsert{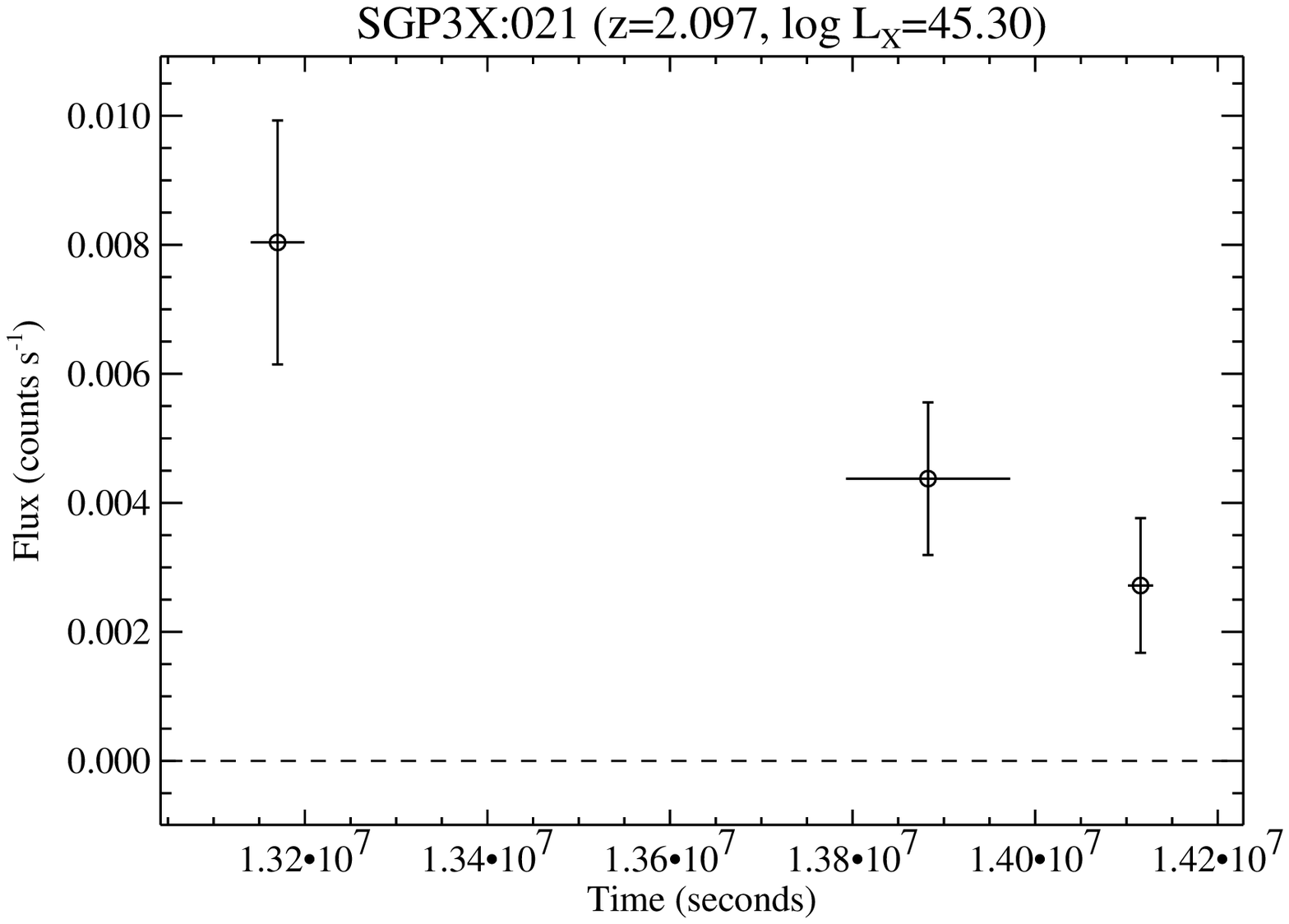}{0.0pt}}
\centerline{\epsfxsize=8.0 truecm \figinsert{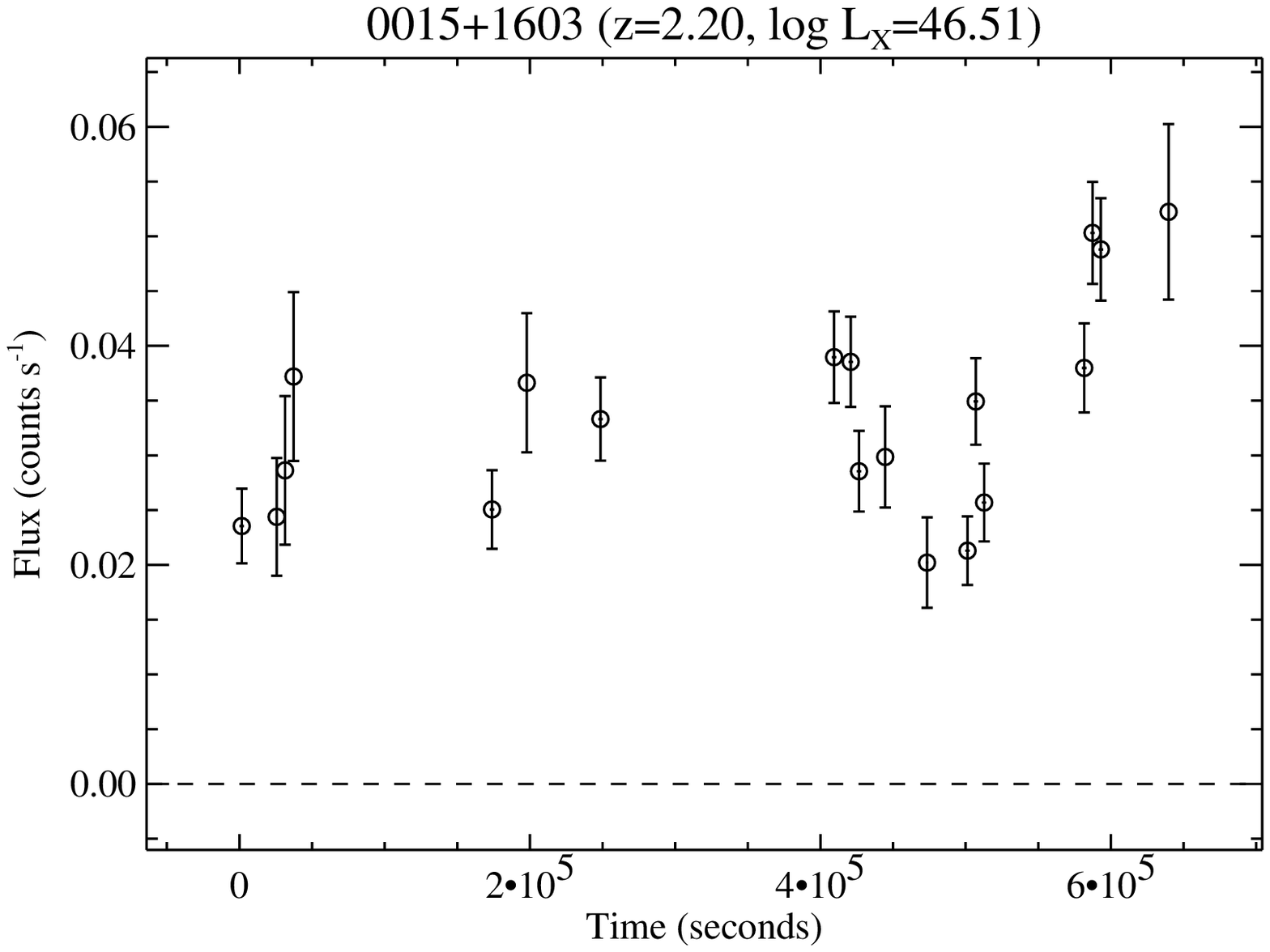}{0.0pt}}
\centerline{\epsfxsize=8.0 truecm \figinsert{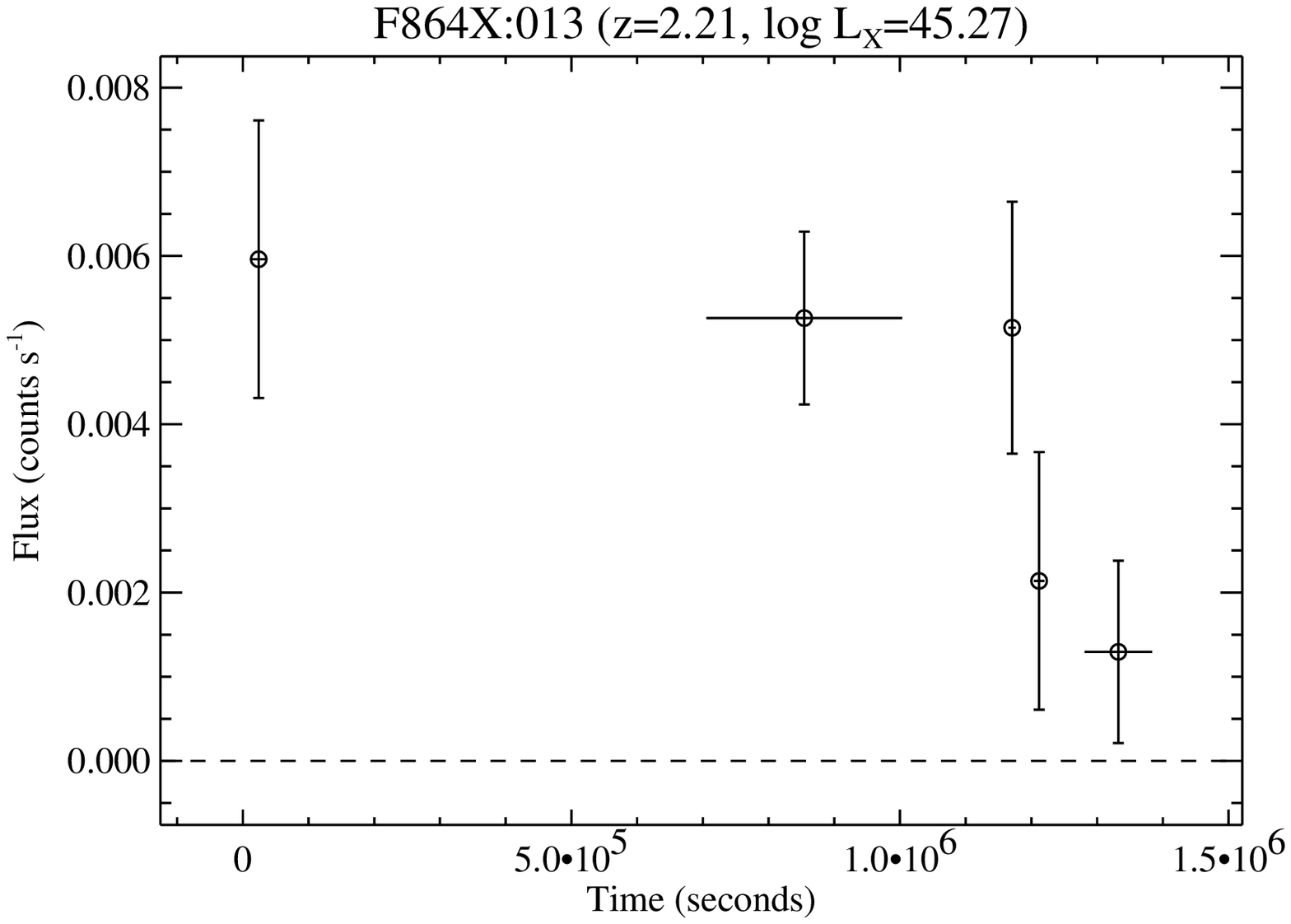}{0.0pt}}
\centerline{\epsfxsize=8.0 truecm \figinsert{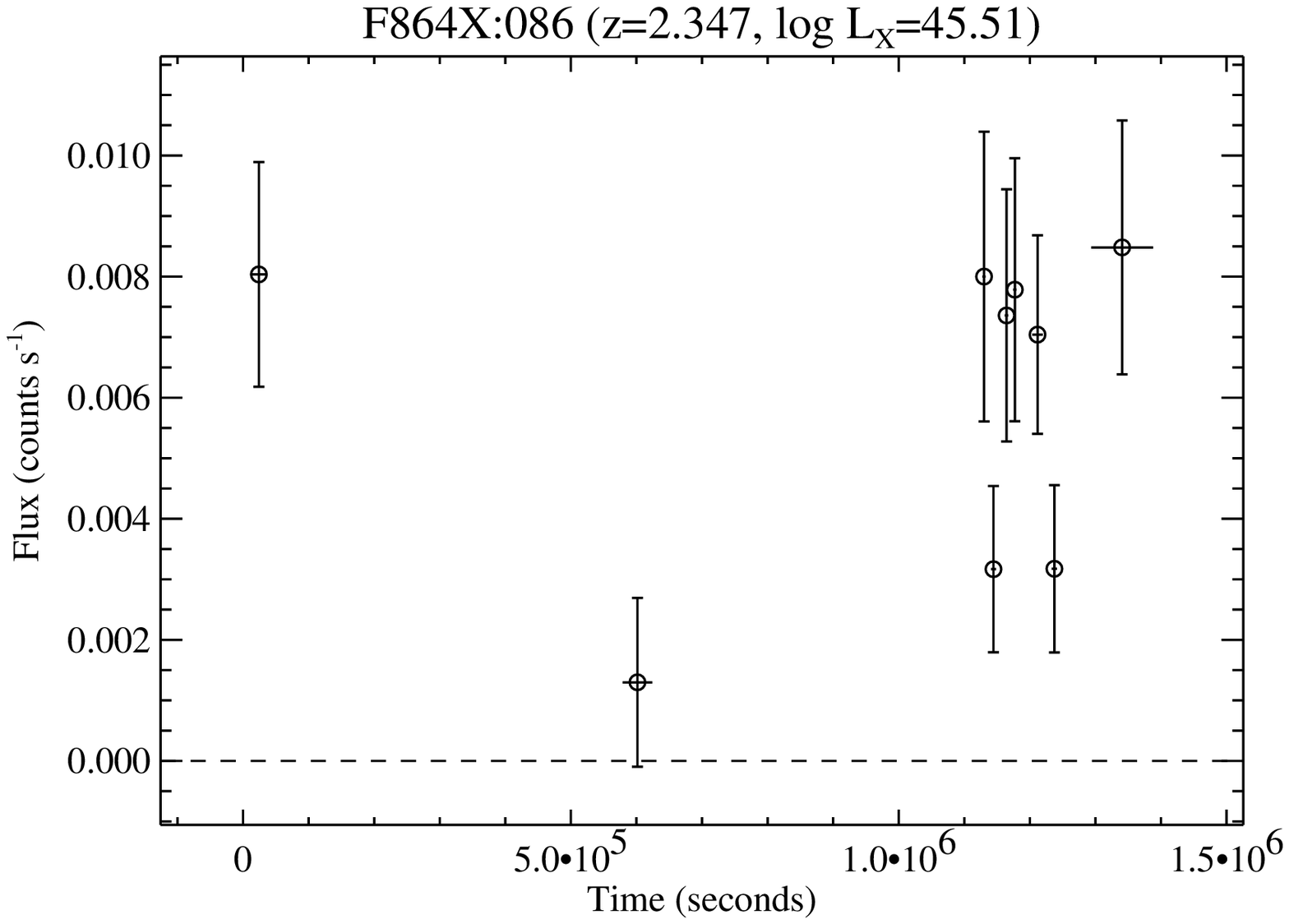}{0.0pt}}
\end{figure}
\addtocounter{figure}{-1}
\begin{figure}
\caption{High-z variable QSOs.}
\centerline{\epsfxsize=8.0 truecm \figinsert{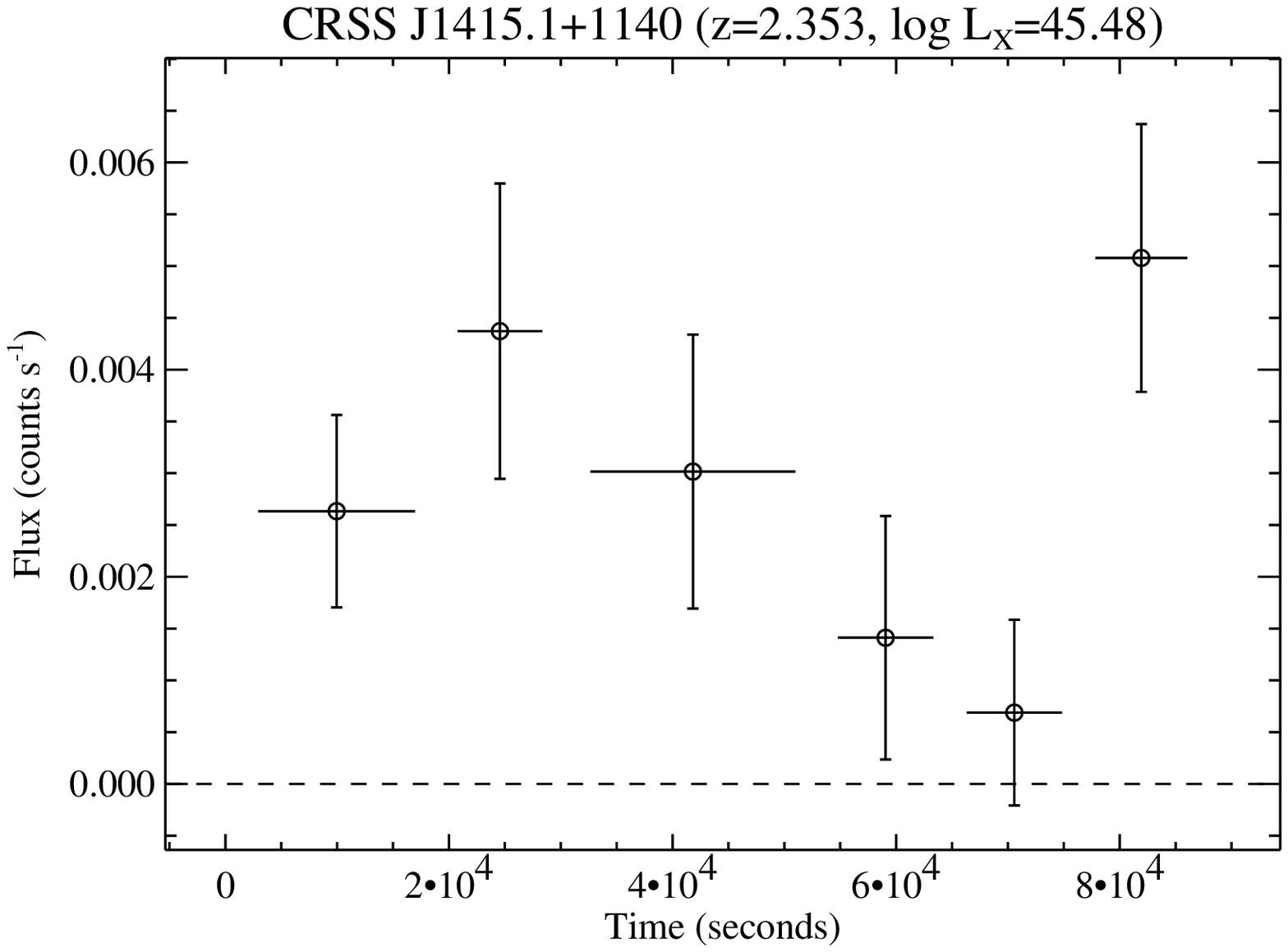}{0.0pt}}
\centerline{\epsfxsize=8.0 truecm \figinsert{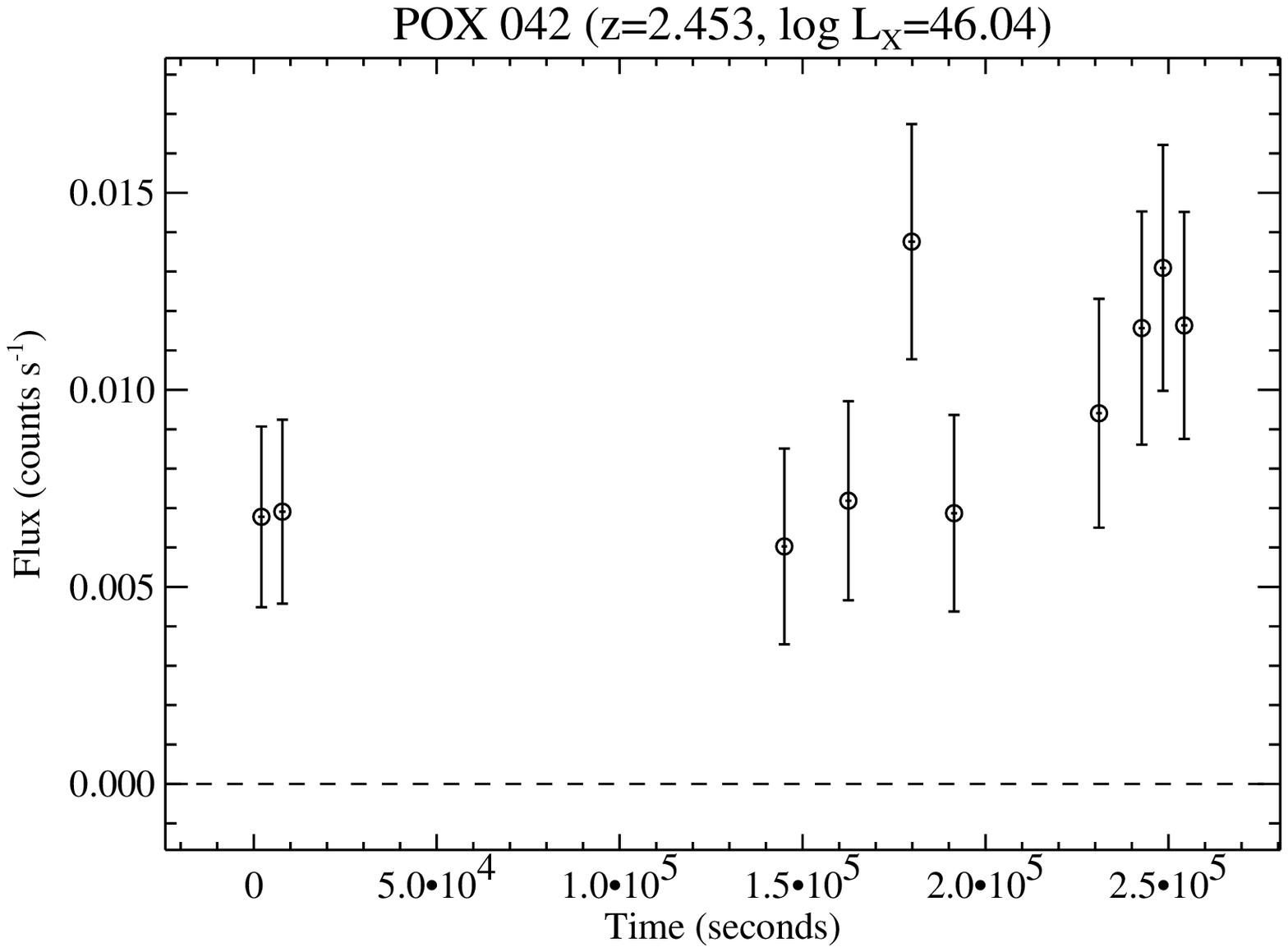}{0.0pt}}
\centerline{\epsfxsize=8.0 truecm \figinsert{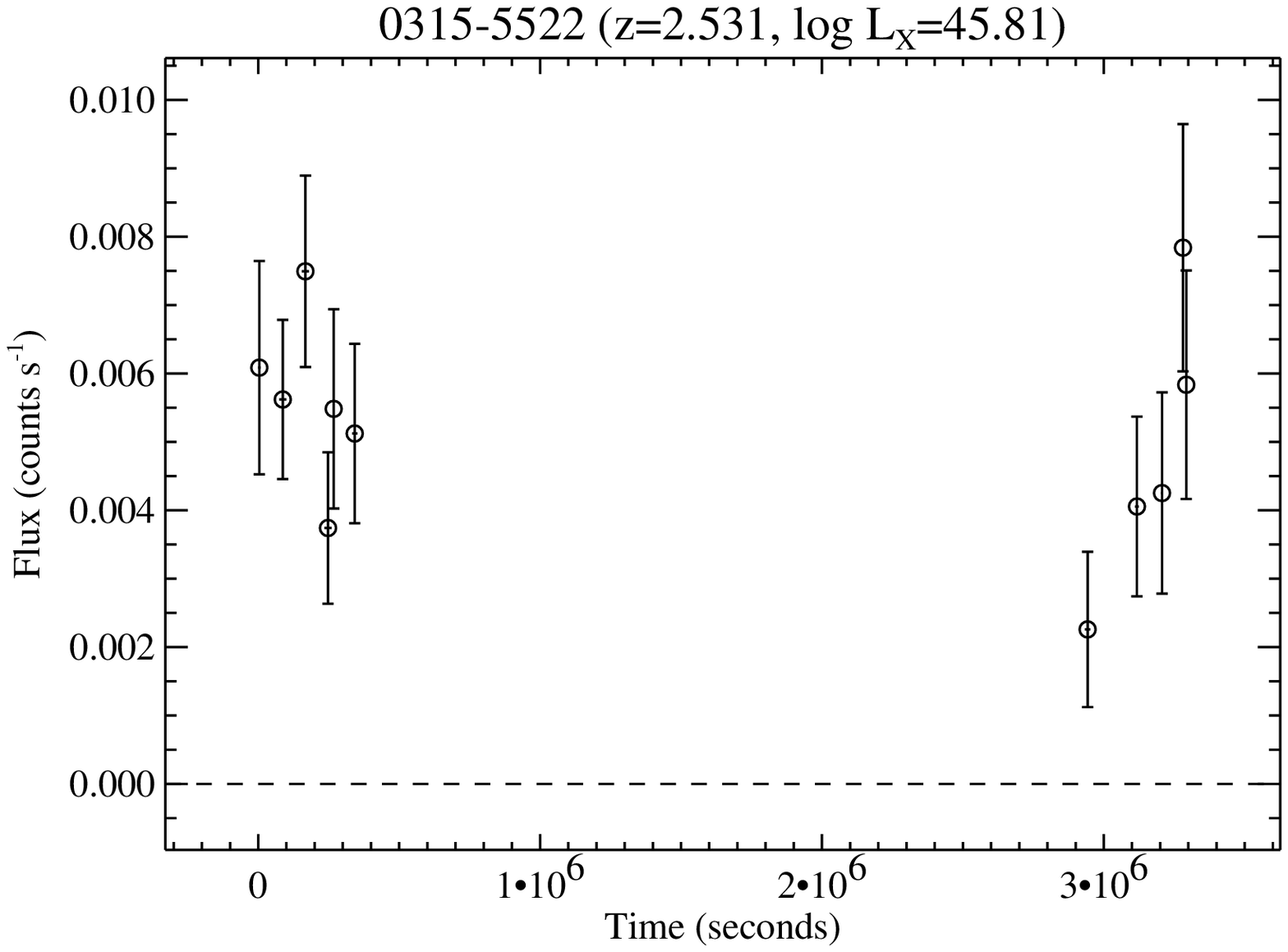}{0.0pt}}
\centerline{\epsfxsize=8.0 truecm \figinsert{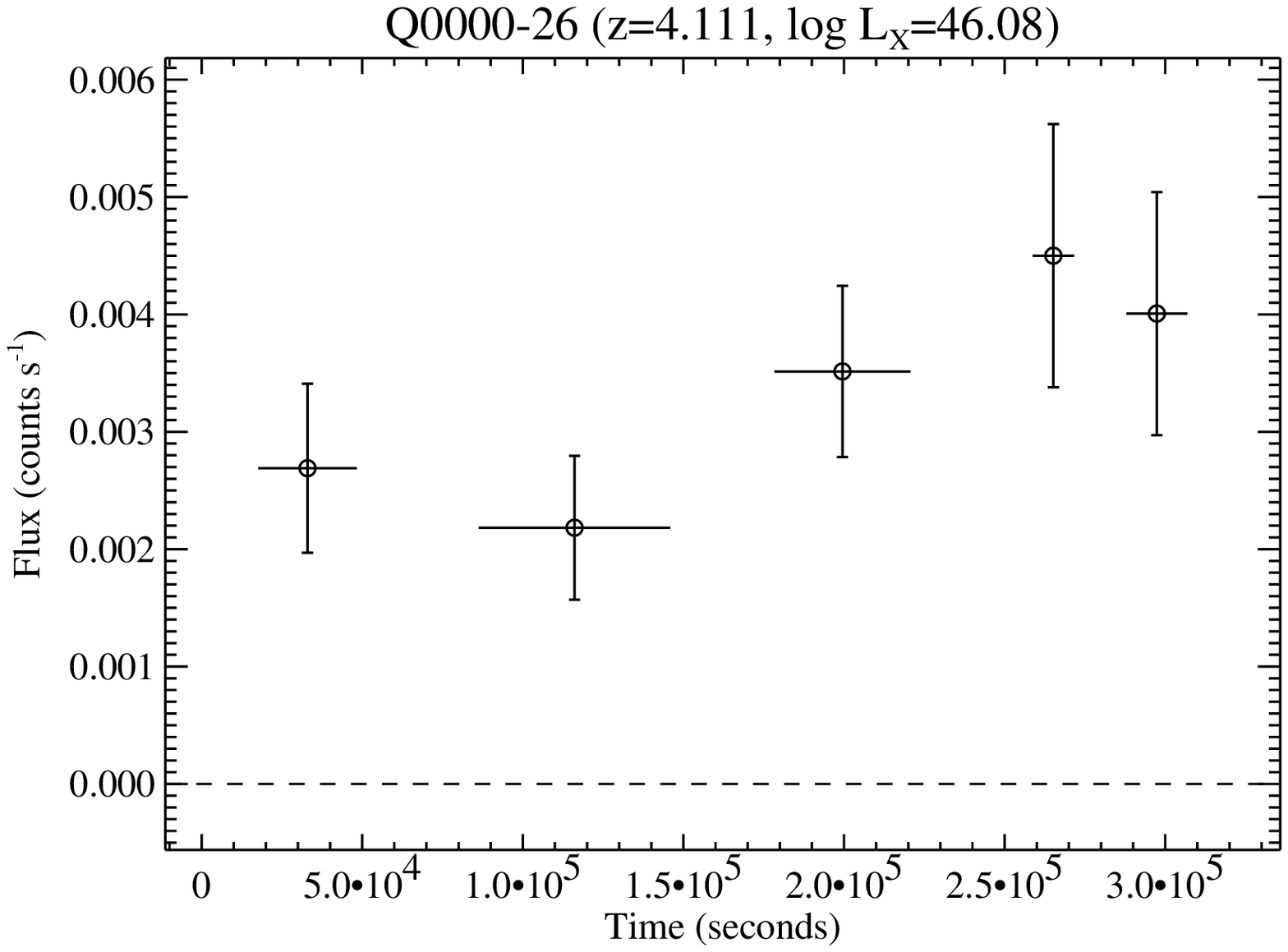}{0.0pt}}
\end{figure}

\section{Implications for Quasar Models}

\subsection{The $\sigma - L_X$ anti-correlation}

The relation between X-ray variability and luminosity, reliably
identified for local AGN, is here seen to continue to redshifts of
$\sim$ 2. However, there is still no definitive explanation for the
anti-correlation. In particular, it is unclear whether variability is
intrinsically linked with luminosity, or whether the link is to a
third parameter that happens to scale with luminosity. If the upturn
in variability observed at high redshifts is real, then an extra
parameter must exist.

A change in measured variability amplitude can occur when the power
spectrum of a quasar light curve is shifted in either amplitude or
time-scale. At present these cannot be distinguished as features in
the power spectrum (such as a turn-over of the power law at low
frequency) have only been identified for the most well studied
AGN (e.g. Edelson \& Nandra 1999, Pounds et al 2001). This leaves us
with a variety of possible explanations for the $\sigma - L_X$
relation, for example:

\begin{itemize}
\item A flaring accretion disk may cause the observed correlation in
two ways. The luminosity may be related to i) the {\em number} of
flares present on the disk (assumed to be all identical), or ii) the
{\em size} of the individual flares.
In the first case, an increase in the number of overlapping flares
would act to `smear out' the observed variability leading to an
amplitude shift in the quasar power spectrum. The second case could
lead to a time-scale effect if flares of longer duration are
responsible for the increased luminosity.
\item X-ray variability time-scale may directly scale with black hole
mass if, for example, the emission occurs at a fixed number of
Schwarzschild radii.
\item Regardless of black hole mass, variability may depend on X-ray
source size, which could in turn depend on luminosity.
\end{itemize}

Recent studies by Ptak et al (1998) show that most low-luminosity
AGN (LLAGN) display very little X-ray variability, in marked contrast
to the relation found by N97. However, these objects
are expected to harbour relatively massive black holes with low
accretion rates. Iwasawa et al (2000) studied the variability of a
dwarf Seyfert (NGC 4395) which is thought to contain a small black hole
($\sim 10^5 M_{\odot}$). They find a large variability amplitude
consistent with an extrapolation of the power-law fit to the $\sigma -
L_X$ relation for the sample of N97. This would
suggest that X-ray variability correlates with black hole mass, rather
than directly with luminosity.

\subsection{Explaining the high-redshift upturn}

Given the possible reasons for the $\sigma - L_X$ anti-correlation, we
can now identify potential causes for a high-redshift upturn in
variability. It is tempting to think this may be simply due to a smaller
typical black hole mass at these early epochs. In practice, this
scenario does not work. Smaller black holes may lead to increased
variability, but they should also lead to lower luminosities. These
QSOs would then just fit an extrapolation of the measured $\sigma -
L_X$ relation. Therefore, we may be seeing an intrinsic change in the
behaviour of these objects. For the same variability seen in local
AGN, these high redshift quasars appear almost an order of magnitude
more luminous. The root of this shift in behaviour could be a change
in the following parameters:

\begin{description}
\item [{\bf Intrinsic change in luminosity (i.e. accretion
efficiency).}] It seems reasonable to assume the
amplitude of variability is linked with a global property of the QSO,
such as black hole mass. The luminosity of the quasar must in some way
be related to the rate of fueling ($\dot{M}$). Seeing a different
$\sigma - L_X$ relation at high redshift then suggests that for a
given black hole mass, the high-z quasars must have a greater rate of
fueling, in other words, they are accreting at a higher fraction of
the Eddington limit. Is it then feasible that $z > 2$ QSOs are
accreting more efficiently than local AGN by almost an order of magnitude?

\item The Eddington ratio for local AGN has been relatively well
constrained. Wandel et al (1999), use reverberation methods to
accurately define the Eddington ratio for local Seyferts. They find
$L/L_{Edd} \approx 0.01-0.3$, though there is a strong trend of the
Eddington ratio to increase with luminosity. The necessary increase in
accretion efficiency, to allow a typical high redshift QSO to be almost
an order of magnitude more luminous, would imply these objects were
accreting very close to the Eddington limit.

\item If this increase in accretion efficiency is real, the cause must lie
with the environmental conditions at these redshifts. It is plausible
that early QSOs enjoyed a more fuel-rich environment.

\item [{\bf Intrinsic change in variability.}] The upturn may also be
characterized as an increase in variability for QSOs of the same
luminosity. Possible scenarios include:
\begin{enumerate}
\item The X-ray emitting region is physically smaller in size, but of
greater intensity, for high-z QSOs. This could occur if the emission
region was located at a smaller number of Schwarzschild radii.
\item In the flaring accretion disk model, there may be a smaller
number of more luminous flares for high-z QSOs. Enhanced magnetic
fields could conceivably provide a mechanism to achieve this.
\item It is possible that some short time-scale variability is caused
by temporary obscuration of the X-ray source. This could be due to
broad-line clouds moving across the line of site, or perhaps rotations
of a warped accretion disk. It is feasible that high-z QSOs may
contain more of this obscuring material, leading to enhanced
variability over these time-scales.
\end{enumerate}

\item [{\bf Spectral sampling.}] The ROSAT band of 0.1 - 2.4 keV used
for the variability analysis, will sample higher rest frame energies
as redshift increases. For a QSO at a redshift of 2.5, variability
will actually be measured in the rest frame band of 0.35 - 8.4
keV. The increased
variability observed at high redshift could therefore be due to
intrinsic spectral variability. This would require QSOs to be more
variable at harder energies indicating a different source for these 
X-rays. However, N97 showed there was a strong
correlation between hard band (2-10keV) and soft band (0.5-2 keV)
variability in local AGN. Where the correlation breaks down they find
greater variability in the soft band. 

\item [{\bf Emergence of a new population.}] An increase in mean AGN
variability would be observed if a subset of highly variable AGN
(such as Narrow-line Seyfert 1s) were more prevalent at high
redshift. Based on the optical spectra however, we find no evidence
that this is due to the emergence of a `narrow-line' population (see
section 6).
\end{description}

Given the low signal-to-noise of this data it is difficult to
constrain these models beyond the general descriptions given here. A
full power spectrum analysis is required to fully characterize and
confirm this trend. This may be possible with long observations
using XMM-Newton, XEUS or Constellation-X.

\section{Conclusions}

We have measured the amplitude of short-term X-ray variability of 156
radio quiet quasars taken from the ROSAT PSPC archive over a redshift
range of 0.08 - 4.11. In order to identify trends with luminosity and
redshift we have combined light curves into ensembles. For QSOs out to
redshift $\sim$ 2 we find the amplitude of variability decreases with
luminosity as $\sigma \propto L_{X}(0.1-2.4$ keV$)^{-\beta}$ with $\beta
= 0.27 \pm 0.05$. This is comparable to the relation found for local AGN. The
behaviour of QSO variability amplitude with redshift is approximately
flat out to $z \sim 2$, although there is some suggestion of a minimum
at $z \sim 1.7$. Beyond redshift $\sim$ 2 there is some evidence for an
increase in QSO X-ray variability. The hypothesis that these quasars
observe the local $\sigma - L_{X}$ relation found by Nandra et al
(1997), is marginally rejected. We explore the
possible reasons for this upturn. If the amplitude of X-ray
variability is linked to black hole mass, this would imply nearly an order of
magnitude increase in accretion efficiency for typical high-redshift
QSOs.

\bigskip

\noindent The data used to produce any of the figures in this paper may be
obtained by emailing James Manners at jcm@roe.ac.uk.

\section*{ACKNOWLEDGMENTS}

This research has made use of data obtained from the Leicester
Database and Archive Service at the Department of Physics and
Astronomy, Leicester University, UK. James Manners acknowledges the
support of a PPARC Studentship.

\end{document}

%% file: epsf_mn.tex
\newread\epsffilein    
\newif\ifepsffileok    
\newif\ifepsfbbfound   
\newif\ifepsfverbose   
\newdimen\epsfxsize    
\newdimen\epsfysize    
\newdimen\epsftsize    
\newdimen\epsfrsize    
\newdimen\epsftmp      
\newdimen\pspoints     
\def\epsfbox#1{\global\def\epsfllx{72}\global\def\epsflly{72}%
   \global\def\epsfurx{540}\global\def\epsfury{720}%
   \def\lbracket{[}\def\testit{#1}\ifx\testit\lbracket
   \let\next=\epsfgetlitbb\else\let\next=\epsfnormal\fi\next{#1}}%
\def\epsfgetlitbb#1#2 #3 #4 #5]#6{\epsfgrab #2 #3 #4 #5 .\\%
   \epsfsetgraph{#6}}%
\def\epsfnormal#1{\epsfgetbb{#1}\epsfsetgraph{#1}}%
\def\epsfgetbb#1{%
\openin\epsffilein=#1
\ifeof\epsffilein\errmessage{I couldn't open #1, will ignore it}\else
   {\epsffileoktrue \chardef\other=12
    \def\do##1{\catcode`##1=\other}\dospecials \catcode`\ =10
    \loop
       \read\epsffilein to \epsffileline
       \ifeof\epsffilein\epsffileokfalse\else
          \expandafter\epsfaux\epsffileline:. \\%
       \fi
   \ifepsffileok\repeat
   \ifepsfbbfound\else
    \ifepsfverbose\message{No bounding box comment in #1; using defaults}\fi\fi
   }\closein\epsffilein\fi}%
\def\epsfsetgraph#1{%
   \epsfrsize=\epsfury\pspoints
   \advance\epsfrsize by-\epsflly\pspoints
   \epsftsize=\epsfurx\pspoints
   \advance\epsftsize by-\epsfllx\pspoints
   \epsfxsize\epsfsize\epsftsize\epsfrsize
   \ifnum\epsfxsize=0 \ifnum\epsfysize=0
      \epsfxsize=\epsftsize \epsfysize=\epsfrsize
     \else\epsftmp=\epsftsize \divide\epsftmp\epsfrsize
       \epsfxsize=\epsfysize \multiply\epsfxsize\epsftmp
       \multiply\epsftmp\epsfrsize \advance\epsftsize-\epsftmp
       \epsftmp=\epsfysize
       \loop \advance\epsftsize\epsftsize \divide\epsftmp 2
       \ifnum\epsftmp>0
          \ifnum\epsftsize<\epsfrsize\else
             \advance\epsftsize-\epsfrsize \advance\epsfxsize\epsftmp \fi
       \repeat
     \fi
   \else\epsftmp=\epsfrsize \divide\epsftmp\epsftsize
     \epsfysize=\epsfxsize \multiply\epsfysize\epsftmp   
     \multiply\epsftmp\epsftsize \advance\epsfrsize-\epsftmp
     \epsftmp=\epsfxsize
     \loop \advance\epsfrsize\epsfrsize \divide\epsftmp 2
     \ifnum\epsftmp>0
        \ifnum\epsfrsize<\epsftsize\else
           \advance\epsfrsize-\epsftsize \advance\epsfysize\epsftmp \fi
     \repeat     
   \fi
   \ifepsfverbose\message{#1: width=\the\epsfxsize, height=\the\epsfysize}\fi
   \epsftmp=10\epsfxsize \divide\epsftmp\pspoints
   \newcount\figskipcount
      \message{#1 \the\epsfysize  }
   \vbox to\epsfysize{\vfil\hbox to\epsfxsize{%
      \includegraphics{#1}%
      \hfil}}%
\epsfxsize=0pt\epsfysize=0pt}%

{\catcode`\%=12 \global\let\epsfpercent=
\long\def\epsfaux#1#2:#3\\{\ifx#1\epsfpercent
   \def\testit{#2}\ifx\testit\epsfbblit
      \epsfgrab #3 . . . \\%
      \epsffileokfalse
      \global\epsfbbfoundtrue
   \fi\else\ifx#1\par\else\epsffileokfalse\fi\fi}%
\def\epsfgrab #1 #2 #3 #4 #5\\{%
   \global\def\epsfllx{#1}\ifx\epsfllx\empty
      \epsfgrab #2 #3 #4 #5 .\\\else
   \global\def\epsflly{#2}%
   \global\def\epsfurx{#3}\global\def\epsfury{#4}\fi}%
\def\epsfsize#1#2{\epsfxsize}